\documentclass[review]{elsarticle}

\usepackage{tikz}
\usepackage{amsmath}
\usepackage{subfigure}
\usepackage{geometry}
\usepackage{lineno,hyperref}
\modulolinenumbers[5]

\journal{Elsevier}

\begin{document}

\begin{frontmatter}

\title{A unified construction of all-speed HLL-type schemes for hypersonic heating computations}

\author[mymainaddress]{Wenjia Xie\corref{mycorrespondingauthor}}
\cortext[mycorrespondingauthor]{Corresponding author}
\ead{xiewenjia@nudt.edu.cn}

\author[mymainaddress]{Ye Zhang}
%\author[mymainaddress]{Hang Yu}

\author[mymainaddress]{Zhengyu Tian}
\author[mymainaddress]{Hua Li}

\address[mymainaddress]{College of Aerospace Science and Engineering, National University of Defense Technology, Hunan 410073, China}

\begin{abstract}
In this paper, a unified framework to develop all-speed HLL-type schemes for hypersonic heating computations is constructed. Such a unified construction method combines two effective improving techniques: a shock robustness improvement and a low-Mach number fix. It is implemented by properly modifying the approximate solutions of the local Riemann problem in the HLL framework, resulting in two all-speed HLL-type schemes, namely ASHLLC and ASHLLEM solvers. Results from both numerical analysis and experiments demonstrate that the newly proposed schemes not only preserve desirable properties of their original versions, but are also able to provide accurate and robust solutions for complex flows ranging from low-Mach number incompressible to hypersonic compressible regimes. Thus, both the ASHLLC and ASHLLEM schemes can be used as reliable methods for hypersonic heating computations.
\end{abstract}


\end{frontmatter}

%\linenumbers

\section{Introduction}
\label{S:1}
Hypersonic aeroheating prediction is one of the key technologies for the development of hypersonic vehicle. With the development of numerical methods and the computer technology, computational fluid dynamics (CFD) has increasingly become one of the most popular approaches to predict hypersonic aeroheating. However, despite the great progress made in the past decades, it is still challenging for the current CFD methods to reliably predict hypersonic heating problems \cite{candler2009current,gnoffo2010challenges}.

The challenge of correctly predicting hypersonic heat transfer stems from a variety of factors, such as physical models, flow conditions, numerical methods, mesh topology and so on. One of the primary difficulties is robustly capturing of strong shock waves in hypersonic flows \cite{kitamura2010evaluation}, because the computation of hypersonic heating transfer is strongly sensitive to shock anomalies, such as the carbuncle phenomenon and the post-shock oscillations. Numerical errors which are induced by shock anomalies will propagate to the downstream boundary layer, resulting into inaccurate predictions of temperature gradients at the vehicle surface. Currently, second-order finite volume methods combined with limiters are commonly used for hypersonic flow computations due to their conservative property and robust performance. Previous researches and practical applications have demonstrated that these finite volume methods succeed to reliably predict heat transfer in a variety of hypersonic flow configurations in which the mesh is specifically aligned with strong shocks. Recent studies \cite{barter2010shock,burgess2012computing,brazell20133d,ching2019shock} also show active efforts to develop high-order methods for hypersonic heating computations. Whereas, even for the low-order finite volume methods, current numerical methods cannot always provide good heating predictions, particularly on non-shock-aligned meshes. To resolve strong shock waves robustly, additional numerical dissipation is required to control perturbation errors across the bow shock, but meanwhile such dissipation should be minimized in the stagnation region and the boundary layer to accurately resolve the temperature gradients. Thus, the dissipative property of shock-capturing schemes used for hypersonic heating should be carefully balanced between accuracy and robustness. Furthermore, recent studies by Kitamura et al. \cite{kitamura2010evaluation,kitamura2013further,kitamura2013towards} have demonstrated that in addition to inviscid flux functions, the choice of reconstructed variables and the limiter along with associated parameters also exert important influence on heat transfer predictions, particularly in the case where the alignment of the computational mesh with the shock wave is poor.

In the past decades, great efforts have been made to develop reliable numerical schemes for real hypersonic applications with complex geometries. In hypersonic flow regimes, it is still challenging for numerical schemes to predict shock wave robustly. Shock anomalies such as carbuncle phenomenon and post-shock oscillations are often encountered due to different flow conditions, mesh systems and numerical methods. A variety of schemes have been proposed to improve performance of schemes in the hypersonic flow regime. One of the most prominent methods is the advection upstream splitting method (AUSM), originally developed by Liou and Steffen \cite{liou1993new}, and its variant AUSM-family schemes \cite{liou1996sequel,wada1997accurate,kim2001methods} are simple, accurate and robust for hypersonic flow computations. Thus, they have been widely used as one of the most common methods for practical compressible flow applications. Further extensions of AUSM-family schemes to all-speed schemes that can be applied from low Mach number to high Mach number flows are also developed by Edwards \cite{edwards2001towards} and Liou \cite{liou2006sequel}. In recent years, Kitamura and Shima develop a series of all-speed shock-capturing schemes based on AUSM family schemes, named SLAU (Simple Low-dissipative AUSM) \cite{shima2011parameter}, SLAU2 \cite{kitamura2013towards} and SD (Shock Detecting)-SLAU \cite{Shima2013}, which have been successfully applied for flows involving both low and high Mach number regions. Numerical experiments \cite{kitamura2013towards,kitamura2013further,qu2018study} demonstrate that these all-speed schemes have good performance in hypersonic heating computations. Similar improved all-speed flux splitting flux functions based on AUSM-family schemes are also developed by Qu et al. \cite{qu2017improvement,qu2018new,Qu2019}. Their proposed AUSM-type schemes have also shown great potential in computing hypersonic heating problems. Furthermore, in a comparative study \cite{qu2016investigation}, the same authors argue that a high level of accuracy at low speeds is beneficial to the hypersonic heating computations.

Another approach commonly used for compressible flow computations is the flux difference splitting (FDS) method, which can be generally classified into two groups: complete and incomplete Riemann solvers. The popular Roe scheme \cite{roe1981approximate}, Osher scheme \cite{osher1982upwind}, HLLEM scheme \cite{Einfeldt1988,Einfeldt1991} and HLLC scheme \cite{Toro1994,toro2013riemann} belong to the complete solver. These schemes have the same wave structure as the exact Riemann solver and own minimal dissipation on contact and shear waves. Thus, they are suitable for computing viscous flows. However, such low dissipative property is much more likely to trigger shock instabilities, which limit their applications in hypersonic flow regime. The incomplete Riemann solvers include the HLL scheme \cite{harten1983upstream}, the HLLE scheme \cite{Einfeldt1991} and the Rusanov scheme \cite{rusanov1961calculation}. These numerical methods lead to excessive dissipation of contact waves, thus they are generally endowed with high resistance to shock instabilities. Whereas, excessive dissipation introduced by these solvers reduce the resolution of boundary layers, which is crucial for accurate prediction of surface heat transfer. Thus, in order to develop a reliable method for hypersonic flows, several hybrid techniques \cite{shen2016robust,xie2017numerical,zhang2017robust,simonnumerical} are proposed to improve the shock robustness of complete solvers by  hybridizing the incomplete solvers. These hybrid schemes have demonstrated good performance of the computation of hypersonic flows, but their performance in hypersonic heating computations is seldom assessed. Furthermore, these solvers developed for compressible flows usually cannot maintain the accuracy in the low-speed flow regime. In reference \cite{xie2017numerical}, we are able to clarify that the numerical shock instability is strongly related to perturbation errors and their propagation in the vicinity of strong shocks. Based on these findings, an effective approach of suppressing shock instabilities is proposed to improve shock robustness of the Roe scheme \cite{xie2019towards} and the HLLC scheme \cite{xie2019accurate}. Such an improved approach is capable of resolving strong shock waves without introducing additional shear viscosity in hypersonic flow regime, demonstrating favorable potential for hypersonic heating computations. In the current study, we make further efforts to develop accurate and reliable shock-capturing methods for compressible flows ranging from low Mach number to high Mach number regimes, and assess their performance in hypersonic heating applications. To this end, a unified framework to develop all-speed shock-capturing schemes for hypersonic heating computations is constructed. These schemes are built on top of famous HLL-type schemes, i.e., HLLEM and HLLC, which are able to capture discontinuities sharply and enjoy desirable properties such as entropy satisfaction and positivity preservation. By combing an efficient improvement for the shock instability and a low-Mach correction, the accuracy and robustness of both methods are improved significantly in the same HLL-type framework. Numerical analysis and experiments demonstrate that the proposed all-speed solvers can not only provide accurate results for all flow regimes, but also solve strong shock waves robustly even on enlonged non-shock-aligned meshes. All of these favorable properties suggest that the current all-speed HLL-type schemes are promising to be widely used to accurately and efficiently simulate various complex flows including the hypersonic heating problem.

The outline of the rest of this paper is as follows. In section 2, governing equations of compressible flows and their related finite volume discretization are presented. Three classical HLL-type schemes are also reviewed in the same section. In section 3, a unified framework for constructing all-speed HLL-type schemes are presented. In section 4, two important properties of the current modified flux functions are clarified by both the analysis method and numerical experiments. The accuracy and robustness of the proposed methods for all-speed flow problems especially the hypersonic heating problem are tested in section 5. Section 6 contains conclusions and an outlook to future developments.

\section{Governing equations and numerical method}
\label{S:2}

We consider a compressible flow governed by the three-dimensional Navier-Stokes equations written in integral form as
\begin{equation}\label{2.1}
    \frac{\partial}{\partial t} \int_{\Omega} \mathbf{U} {\rm{d}}{\Omega} + \oint_{\partial \Omega} ( {\mathbf{F}}_c - {\mathbf{F}}_v) {\rm{d}}S = 0,
\end{equation}
where $\partial \Omega$ denote boundaries of the control volume $\Omega$. The state vector and convective flux vector are defined as
\begin{equation}\label{2.2}
	\mathbf{U} = \left[ {\begin{array}{{c}}
	\rho \\
	{\rho u}\\
	{\rho v}\\
	{\rho w}\\
	{\rho e}
	\end{array}} \right],
	\quad
	{\mathbf{F}}_c = \left[ {\begin{array}{{c}}
	\rho q \\
	{\rho u q + p n_x}\\
	{\rho v q + p n_y}\\
	{\rho w q + p n_z}\\	
	{\left(\rho e+p\right)q}
	\end{array}} \right],
\end{equation}
where $\rho$, $e$, and $p$ represent density, specific total energy and pressure respectively, and ${\bf{u}} = \left( {u,v,w} \right)$ is the flow velocity. The directed velocity, $q=un_x+vn_y+wn_z$, is the component of velocity acting in the $\mathbf{n}$ direction, where $\mathbf{n}={\left[ {{n_x},\;{n_y},\;{n_z}} \right]^T}$ is the outward unit vector normal to the surface element ${\rm{d}}S$. The definition of the vector of viscous fluxes ${\mathbf{F}}_v$ can be referred to \cite{Blazek}. To close the set of equations, the idea-gas equation of state is used, i.e. the pressure is given by $p=(\gamma -1) \rho e$ with a constant ratio of specific heats $\gamma$.

\subsection{Finite volume method}
\label{S:2.1}
To solve the Navier-Stokes equations numerically, the Godunov's approach for finite volumes is applied. The semi-discrete finite volume scheme over a particular control volume ${\Omega _i}$  can be written as
\begin{equation}\label{2.4}
\frac{{{\rm{d}}{{\bf{U}}_i}}}{{{\rm{d}}t}} + \frac{1}{{\left| {{\Omega _i}} \right|}}\sum\limits_{{\Gamma _{ij}} \subset \partial {\Omega _i}} {\left| {{\Gamma _{ij}}} \right|}  \left({ {\bf{F}}_c-{\bf{F}}_v}\right)_{ij} = 0.
\end{equation}
In the above expression, ${{\bf{U}}_i}$ is the cell average of ${\bf{U}}$ on ${\Omega _i}$ , $\left| {{\Omega _i}} \right|$  denotes the volume of  ${\Omega _i}$. ${\Gamma _{ij}}$ denotes the common edge of two neighboring cells ${\Omega _i}$ and ${\Omega _j}$, ${{\bf{n}}_{ij}}$ represents the unit vector normal to ${\Gamma _{ij}}$ pointing from ${\Omega _i}$ to ${\Omega _j}$ and $|{\Gamma _{ij}}|$ is the length of face ${\Gamma _{ij}}$. The flux ${{\bf{F}}_{c,ij}}$ is the calculated numerical convective flux that is supposed to be constant along the individual face  ${\Gamma _{ij}}$. It is determined dimension-by-dimension from an approximate Riemann solver presented in the following sections. The numerical viscous flux ${{\bf{F}}_{v,ij}}$ at the cell interface is approximated using central difference.

\subsection{Approximate Riemann solvers}
\label{S:2.2}
\begin{figure}[htbp]
  \centering
  \includegraphics[width=0.70\textwidth]{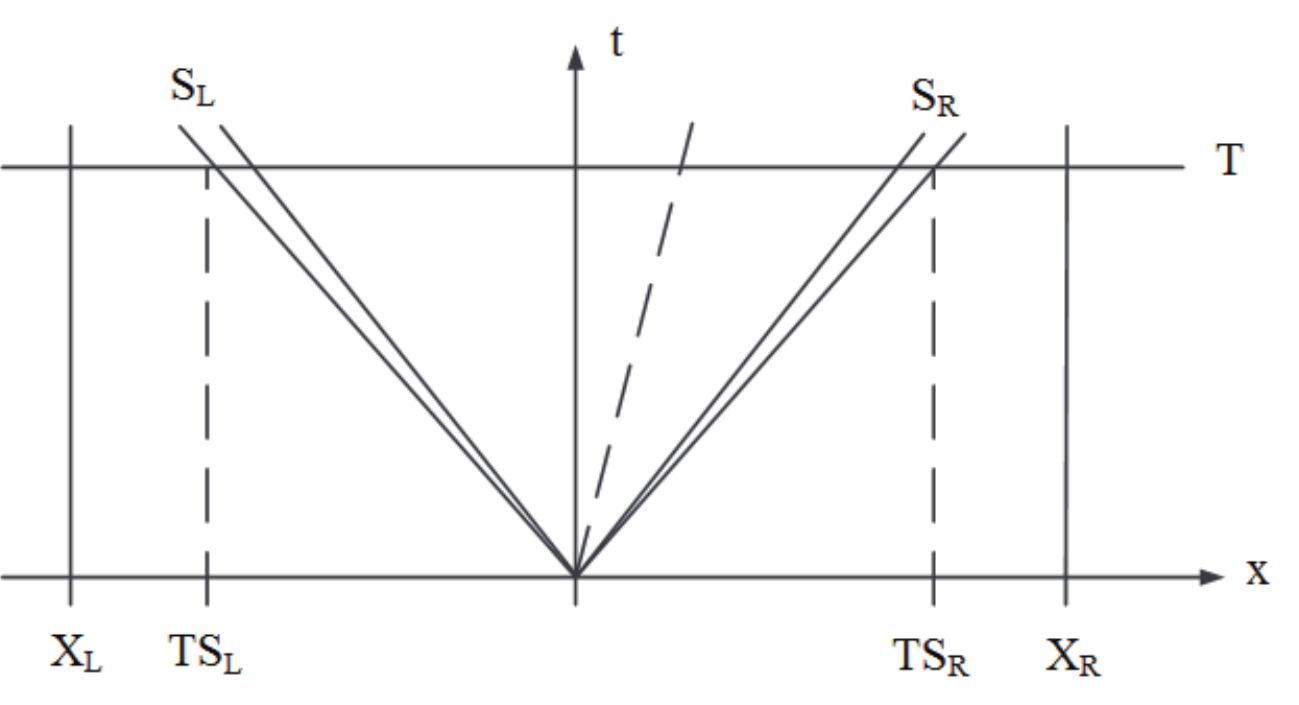}
  \caption{Control volume $[x_L, x_R]$ $\times$ $[0, T]$ on $x-t$ plane. $S_L$ and $S_R$ are the fastest signal velocities arising from the solution of the Riemann problem \cite{toro2013riemann}. }
  \label{fig1}
\end{figure}

The key step of the Godunov's approach is to solve the following Riemann problem of the one-dimensional Euler equations,
\begin{equation}\label{RM1}
  \frac{\partial{\bf{U}}}{\partial t}+\frac{\partial{\bf F}_c}{\partial x}=0,
\end{equation}
\begin{equation}\label{RM2}
{\bf{U}}\left( {x,0} \right) =
	\begin{cases}
	{\bf U}_L       & \text{$x < 0$},    \\
	{\bf U}_R       & \text{$x > 0$}.
	\end{cases}
\end{equation}
Here, only plane waves parallel to the y-axis are assumed without loss of generality. In the current study, the classical HLLC and HLLEM approximate Riemann solvers, built on top of the HLL scheme \cite{harten1983upstream}, are used to approximate the solution of system (\ref{RM1}) and (\ref{RM2}). Both methods are well-known to be entropy stable, positivity preserving and robust for compressible flow computations. However, in hypersonic flows where strong shocks exist, these methods usually encounter shock instability problems, such as the carbuncle phenomenon. Moreover, like other shock-capturing schemes designed for compressible flows, these HLL-type solvers also produce excess numerical dissipation in incompressible regions. These deficiencies limit their applications on simulating complex flows across a wide range of Mach numbers, in terms of accuracy and robustness. To facilitate further discussion, a brief description of these HLL-type schemes are presented in the following sections. In Fig. \ref{fig1}, the complete wave structure arising from the exact solution of the Riemann problem is presented in the control volume $[x_L, x_R]$ $\times$ $[0, T]$.

\subsubsection{HLL solver}
\label{S:2.2.1}
The HLL Riemann solver proposed by Harten, Lax, and van Leer \cite{harten1983upstream} is one of the most reliable approaches to solve the Riemann problem approximately. Different from the exact Riemann solution with a large amount of detail, the HLL solver assumes that the solution consists of three constant states, separated by two waves propagating at speeds of $S_L$ and $S_R$, i.e.,
\begin{equation}\label{eq2.2.1_1}
	{\widetilde {\bf{U}}} (x/t)=
	\begin{cases}
	{\bf U}_L       & \text{$x/t \le S_L$}     \\
	{\bf U}^{hll}       & \text{$S_L \le x/t \le S_R$}  \\
	{\bf U}_R       & \text{$S_R \le x/t$}
	\end{cases},
\end{equation}
where the approximate intermediate state ${\bf{U}}^{hll}$ is defined to satisfy the following consistency condition,
\begin{equation}\label{eq2.2.1_2}
	\int_{x_L}^{x_R} {\bf{U}}(x,T) \;dx = {x_R}{{\bf{U}}_R} - {x_L}{{\bf{U}}_L} + T\left( {{{\bf{F}}_L} - {{\bf{F}}_R}} \right).
\end{equation}
Such a constant state can be denoted by
\begin{equation}\label{eq2.2.1_3}
{{\bf{U}}^{hll}} = \frac{{{S_R}{{\bf{U}}_R} - {S_L}{{\bf{U}}_L} + {{\bf{F}}_L} - {{\bf{F}}_R}}}{{{S_R} - {S_L}}}.
\end{equation}
With the approximate solution (\ref{eq2.2.1_1}) and the definition of ${{\bf{U}}^{hll}}$ in (\ref{eq2.2.1_3}), the corresponding interface flux can be obtained by evaluating the integral form of the conservation laws on the control volume $\left[ {0,{x_R}} \right] \times \left[ {0,T} \right]$, that is,
\begin{equation}\label{eq2.2.1_4}
	{\bf{F}}_{hll}=
	\begin{cases}
	{\bf F}_L       & \text{$S_L \ge 0$}               \\
	\frac{{{S_R}{\bf{F}}_L - {S_L}{\bf{F}}_R}}{{{S_R} - {S_L}}} + \frac{{{S_L}{S_R}}}{{{S_R} - {S_L}}}\left( {{{\bf{U}}_R} - {{\bf{U}}_L}} \right)       & \text{$S_L \le 0 \le S_R$}    \\
	{\bf F}_R       & \text{$S_R \le 0$}
	\end{cases}.
\end{equation}
To completely determine the interface flux, the wave speeds need to be estimated. Here, we adopt the simple wave speeds estimate proposed by Davis \cite{Davis1988},
\begin{equation}\label{eq2.2.1_5}
	S_L=\min(q_L-a_L,q_R-a_R),\quad S_R=\max(q_L+a_L,q_R+a_R).
\end{equation}
where $a_L$ and $a_R$ are the sound speeds of the left and right states.

\subsubsection{HLLEM solver}
\label{S:2.2.2}
Due to the assumption of a two-wave configuration, the HLL scheme introduces excess dissipation on linear waves. Therefore, it provides poor resolution of physical features such as contact surfaces, shear waves and material interfaces. This deficiency can be resolved by properly recovering the missing contact information in the Riemann solution. The first strategy introduced is due to Einfeldt \cite{Einfeldt1988,Einfeldt1991}, who manages to modify the intermediate state in (\ref{eq2.2.1_1}) through a linear distribution approach. The resulting modified flux called HLLEM approximates the Riemann solution in the following way,
\begin{equation}\label{eq2.2.2_1}
{\widetilde {\bf{U}}}(x/t)=
\begin{cases}
{\bf U}_L               & \text{$x/t \le S_L$}  \\
{\bf U}^{hll}+(x-\overline{q} t) ({\hat{\delta}}^*_2 \hat{\alpha}_{2} \widehat{{\bf R}}_{2} + {\hat{\delta}}^*_3 \hat{\alpha}_{3} \widehat{{\bf R}}_{3} )     & \text{$S_L \le x/t \le S_R$} \\
{\bf U}_R               & \text{$S_R \le x/t$}
\end{cases}
\end{equation}
where $\overline{q}$ denotes the numerical approximation of the velocity at the contact discontinuity, it is defined as a simple arithmetic average of wavespeeds $S_L$ and $S_R$,
\begin{equation}\label{eq2.2.2_2}
	\overline{q}=\frac{S_L+S_R}{2}.
\end{equation}
$\widehat \alpha _2$ and $\widehat {\bf{R}}_2$ represent the wave strength and the right eigenvector corresponding to the entropy wave respectively,
\begin{equation}\label{eq2.2.2_3}
	\widehat \alpha _2 = \Delta \rho  - \frac{{\Delta p}}{{{{\widehat a}^2}}},\qquad
		{\widehat {\bf{R}}_2} = \left[ {\begin{array}{*{20}{c}}
			1\\
			{\widehat u}\\
			{\widehat v}\\
			{\widehat w}\\
			{\left( {{{\widehat u}^2} + {{\widehat v}^2} + {{\widehat w}^2}} \right)/2}
			\end{array}} \right].
\end{equation}
${\widehat \alpha _3\widehat {\bf{R}}_3}$ denotes a reformulated compact term with combined two shear wave components \cite{katate},
\begin{equation}\label{eq2.2.2_4}
{\widehat \alpha _3}{\widehat {\bf{R}}_3} = \widehat \rho \left[ {\begin{array}{*{20}{c}}
		0\\
		{\Delta u - \Delta q{n_x}}\\
		{\Delta v - \Delta q{n_y}}\\
		{\Delta w - \Delta q{n_z}}\\
		{\widehat u\Delta u + \widehat v\Delta v + \widehat w\Delta w - \widehat q\Delta q}
		\end{array}} \right].
\end{equation}
The parameters ${\hat{\delta}}^* _2$ and ${\hat{\delta}}^* _3$ denote anti-diffusion coefficients which play a role in controlling the amount of anti-diffusion in the linear degenerate fields,
\begin{equation}\label{eq2.2.2_5}
	{\hat{\delta}}^*_k=\frac{2}{T(S_R-S_L)} {\hat{\delta}}_k  \quad \text{for} \quad k=2,3
\end{equation}
with
\begin{equation}\label{eq2.2.2_6}
	{\hat{\delta}}_2 = {\hat{\delta}}_3 = \frac{{\widehat a}}{{\widehat a + \left| {\widehat {{q}} } \right|}}.
\end{equation}
One should notice that the Roe's averaged velocity $\widehat q$ instead of $\overline{q}$ is used to calculate ${\hat{\delta}} _2$ and ${\hat{\delta}} _3$, which resolves the stationary contact discontinuity exactly \cite{park2003dissipation}.

The approximate solution (\ref{eq2.2.2_1}) also satisfies the consistency condition in (\ref{eq2.2.1_2}), and the corresponding interface flux can also be obtained by evaluating the integral form of the conservation laws on the control volume $\left[ {0,{x_R}} \right] \times \left[ {0,T} \right]$, that is,
\begin{equation}\label{eq2.2.1_7}
	{\bf{F}}_{hllem}=
	\begin{cases}
	{\bf F}_L      & \text{$S_L \ge 0$}               \\
	\frac{{S_R}{\bf{F}}_L - {S_L}{\bf{F}}_R}{{{S_R} - {S_L}}} + \frac{{{S_L}{S_R}}}{{{S_R} - {S_L}}}\left( {{{\bf{U}}_R} - {{\bf{U}}_L}} - {\hat{\delta}} _2 \hat \alpha _2 \widehat {\bf{R}}_2  - {\hat{\delta}} _3 \hat \alpha _3 \widehat {\bf{R}}_3 \right)       & \text{$S_L \le 0 \le S_R$}    \\
	{\bf F}_R       & \text{$S_R \le 0$}
	\end{cases}.
\end{equation}
Here, the wave-speed estimate (\ref{eq2.2.1_5}) is used to completely define the interface flux. One can observe that the HLLEM scheme will reduce to the HLL solver if the coefficients ${\hat{\delta}} _2$ and ${\hat{\delta}} _3$ are eliminated.

\subsubsection{HLLC solver}
\label{S:2.2.3}
A different approach to restore the missing contact and shear waves in the HLL approach was taken by Toro, Spruce and Speares \cite{Toro1994}. Solution of the Riemann problem is approximated by four constant states separated by three waves emerging from the initial discontinuity at the interface, they are defined as
\begin{equation}\label{eq2.2.3_1}
{\widetilde {\bf{U}}}(x/t)=
\begin{cases}
{\bf{U}}_L,   \quad & \text{$x/t \le S_L$}\\
{\bf{U}}_L^*, \quad & \text{$S_L \le x/t \le S^*$}\\
{\bf{U}}_R^*, \quad & \text{$S^* \le x/t \le S_R$}\\
{\bf{U}}_R,   \quad & \text{$S_R \le x/t$}
\end{cases} ,
\end{equation}
where ${\bf{U}}_L^*$  and ${\bf{U}}_R^*$  represent intermediate states at the left and right sides of the contact discontinuity respectively,
\begin{equation}\label{2.6}
  {\bf{U}}_K^* = \left[ {\rho _K^*, \; \rho _K^*{u_K^*}, \; \rho _K^*{v_K^*}, \; \rho _K^*{w_K^*}, \; \rho _K^*e_K^*} \right],    \quad K=L,R,
\end{equation}
and $S_L$, $S_R$ denote the left and right wave speeds.
The corresponding interface flux, denoted by ${\bf{F}}_{hllc}$, is defined as
\begin{equation}\label{2.7}
{\bf{F}}_{hllc} = \left\{ {\begin{array}{*{20}{c}}
{{{\bf{F}}_L}},\quad&{0 \le {S_L}}\\
{{\bf{F}}_L^*},\quad&{{S_L} \le 0 \le S^*}\\
{{\bf{F}}_R^*},\quad&{S^* \le 0 \le {S_R}}\\
{{{\bf{F}}_R}},\quad&{0 \ge {S_R}}
\end{array}} \right. .
\end{equation}
To determine the intermediate fluxes ${{\bf{F}}_L^*}$ and ${{\bf{F}}_R^*}$, one need to consider the following Rankine-Hugoniot conditions across each of the waves of speeds $S_L$, $S^*$ and $S_R$,
\begin{equation}\label{2.8}
  \begin{split}
{\bf{F}}_L^* & = {{\bf{F}}_L} + {S_L}\left( {{\bf{U}}_L^* - {{\bf{U}}_L}} \right)\\
{\bf{F}}_R^* & = {\bf{F}}_L^* + S^*\left( {{\bf{U}}_R^* - {\bf{U}}_L^*} \right)\\
{{\bf{F}}_R} & = {\bf{F}}_R^* + {S_R}\left( {{{\bf{U}}_R} - {\bf{U}}_R^*} \right)
\end{split}
\end{equation}
By jump conditions (\ref{2.8}), the intermediate states in the star region can be derived as
\begin{equation}\label{2.9}
\begin{split}
{\rho}_K^* & = \frac{\alpha_K}{S_K-S^*}\\
{u}_K^* & = u_K+n_x(S^*-q_K)\\
{v}_K^* & = v_K+n_y(S^*-q_K)\\
{w}_K^* & = w_K+n_z(S^*-q_K)\\
{e}_K^* & =e_K+(S^*-q_K)/(S^*+p_K/{\alpha}_K)
\end{split}
\end{equation}
where the contact velocity and pressure in the star region can be obtained by
\begin{equation}\label{2.10}
\begin{split}
S^* & = \frac{{\alpha}_R q_R -{\alpha}_L q_L+p_L-p_R}{{\alpha}_R-{\alpha}_L}\\
p^* & = \frac{{\alpha}_R p_L - {\alpha}_L p_R - {\alpha}_L {\alpha}_R \left(q_L-q_R\right)}{{\alpha}_R-{\alpha}_L}
\end{split}
\end{equation}
In Eq. (\ref{2.9}) and Eq. (\ref{2.10}), we use the following simple notations that are defined by Shen et al. \cite{Shen2016},
\begin{equation}\label{2.11}
  \alpha_L=\rho_L\left(S_L-q_L\right), \quad \alpha_R=\rho_R(S_R-q_R).
\end{equation}
To complete the HLLC Riemann solver, an algorithm to compute the wave speeds $S_L$ and $S_R$ must be given. Here, we also use the simple estimate presented in (\ref{eq2.2.1_5}).

\section{Unified construction of all-speed HLL-type schemes}
\label{S:3}
As we all know, original contact-resolving HLLEM and HLLC schemes will produce shock anomalies for hypersonic flows, and they also fail to compute accurately flows near the incompressible limit. Such deficiencies limit their applications to simulating broader flow regimes, especially the hypersonic aeroheating problem. In the current study, we focus on these two main deficiencies and develop true all-speed HLL-type fluxes.

\subsection{Shock stabilization by pressure dissipative flux}
\label{S:3.1}
It is shown in a recent work \cite{xie2017numerical} that numerical shock instability is closely related to perturbation errors, which are generated inside the numerical shock structure, and their propagation in the vicinity of shocks. Based on these results, a pressure dissipative flux is proposed to suppress the shock instability problem plaguing the HLLC scheme \cite{xie2019accurate}. This strategy can be traced back to a similar modification used to improve shock robustness of Roe scheme \cite{kim2003cures,xie2019towards}. However, the mechanism of improving shock robustness by the pressure dissipative flux has not yet been fully clarified. In the current study, it will be demonstrated that such a dissipative flux plays a significant role in limiting the propagation of perturbation errors in the vicinity of shocks, and thus improve the robustness of original solvers for resolving strong shocks. In this section, we extend the pressure dissipative flux used in our former work \cite{xie2019accurate} to the HLL-type schemes in the current study and reformulate it in the general HLL-type framework. This allows to develop a general method to improve the shock stability of HLL-type schemes.

The solution of the Riemann problem can be modified by adding a pressure dissipative flux to the approximate solution in the star region,
\begin{equation}\label{eq3.1_1}
	{\widetilde {\bf{U}}}(x/t)=
	\begin{cases}
	{\bf U}_L               & \text{$x/t \le S_L$}  \\
	{\bf U}^{hllem/hllc}+(x-\overline{q} t) ({\hat{\delta}}^*_2 \hat{\alpha}^{p}_{2} \widehat{{\bf R}}_{2} )     & \text{$S_L \le x/t \le S_R$} \\
	{\bf U}_R               & \text{$S_R \le x/t$}
	\end{cases}
\end{equation}
where the intermediate conservative states ${\bf U}^{hllem}$ and ${\bf U}^{hllc}$ are defined following (\ref{eq2.2.2_1}) and (\ref{eq2.2.3_1}) as,
\begin{equation}\label{eq3.1_2}
	{\bf U}^{hllem} = 	{\bf U}^{hll}+(x-\overline{q} t) ({\hat{\delta}}^*_2 \hat{\alpha}_{2} \widehat{{\bf R}}_{2} + {\hat{\delta}}^*_3 \hat{\alpha}_{3} \widehat{{\bf R}}_{3} ) ,\qquad
	{\bf U}^{hllc} = {\bf{U}}_{L/R}^*,
\end{equation}
the modified wave strength $\hat{\alpha}^{p}_{2}$ is obtained by only preserving the pressure difference flux in $\hat{\alpha}_{2}$, that is,
\begin{equation}\label{eq3.1_3}
	\hat{\alpha}^{p}_{2}=\frac{{\Delta p}}{{{{\widehat a}^2}}}.
\end{equation}
One can notice that the modification of the average state in (\ref{eq3.1_1}) still satisfies the consistency condition (\ref{eq2.2.1_2}). Therefore, the Riemann solver (\ref{eq3.1_1}) remains in conservation form. The corresponding numerical flux can be obtained by evaluating the integral form of the conservation laws on the control volume $\left[ {0,{x_R}} \right] \times \left[ {0,T} \right]$, that is,
\begin{equation}\label{eq3.1_4}
		{\bf{F}}=
		\begin{cases}
		{\bf F}_L      & \text{$S_L \ge 0$}               \\
        {\bf F}_{hllem/hllc} + {\bf F}_{p}              & \text{$S_L \le 0 \le S_R$}    \\
		{\bf F}_R      & \text{$S_R \le 0$}
		\end{cases}.	
\end{equation}
where ${\bf F}_{hllem/hllc}$ represent the original numerical fluxes defined in (\ref{eq2.2.1_7}) and (\ref{2.7}), and ${\bf F}_{p}$ denotes the modified flux function,
\begin{equation}
{{\bf{F}}_p} = -\frac{{{S_L}{S_R}}}{{{S_R} - {S_L}}}{\hat{\delta} _2}\frac{{\Delta p}}{{{{\widehat a}^2}}}{\widehat {\bf{R}}_2} .
\end{equation}
It is supposed that the numerical flux function ${{\bf{F}}_p}$ serves the purpose of suppressing possible instabilities at strong shocks, thus it is better to be only activated in the vicinity of shocks. To this end, a shock detection function is used. Thus, the resulting pressure dissipative flux can be written by
\begin{equation}\label{fp}
	{{\bf{F}}_p} = \left(f_p-1\right)\frac{{{S_L}{S_R}}}{{{S_R} - {S_L}}}{\hat{\delta}_2}\frac{{\Delta p}}{{{{\widehat a}^2}}}{\widehat {\bf{R}}_2}
\end{equation}
where the function $f_p$ is defined as
\begin{equation}\label{sd1}
	f = \min {\left( {\frac{{{p_L}}}{{{p_R}}},\frac{{{p_R}}}{{{p_L}}}} \right)^3}
\end{equation}
with
\begin{equation}\label{sd2}
	{f_p} = \mathop {\min }\limits_k \left( {{f_k}} \right) .
\end{equation}
Here, $k$ denotes all the interfaces of the left and right cells. Readers are referred to reference \cite{zhang2017robust} for the detail descriptions. It can be obtained from Eq. (\ref{sd1}) and Eq. (\ref{sd2}) that $f_p$ will approach zero at strong shocks where the difference between left and right pressure is large. Thus, the pressure dissipative function ${{\bf{F}}_p}$ is activated and plays a role in suppressing shock instabilities. In regions of smooth flows, $f_p$ will be approximately equal to one. As a result, ${{\bf{F}}_p}$ is eliminated to avoid possible negative effects on smooth flows.

\subsection{Extension to low Mach number flows}
\label{S:3.3}
The above modified flux functions defined in (\ref{eq3.1_4}) are supposed to compute compressible flows from subsonic to hypersonic accurately and robustly. However, when it comes to low Mach incompressible flow, such methods are known to fail to produce accurate numerical results in the low Mach number limit. Here, a further extension of the current HLL-type schemes to low Mach number flows is presented.

It is theoretically discussed by Thornber et al. \cite{thornber2008improved} that first-order Godunov-type schemes, which use piecewise constant variable extrapolation, will produce an artificially large velocity jump at the cell interfaces for low Mach number flows. Even for higher order methods, the velocity jump normal to the cell interface still exists and contributes to the excess numerical dissipation in the low Mach number region. A common cure for this deficiency is to modify the Riemann solver itself to include correct flow physics of low speed flows. Whereas, to construct a unified all-speed method, a general modification should be pursued. Here, a simple and general modification method proposed by Thornber et al. \cite{thornber2008improved} is applied to the HLL-type schemes (\ref{eq3.1_4}) presented in the above section, that is
\begin{equation}\label{eq3.3.1}
{\bf{u}}_L^*=\frac{(1+z){{\bf{u}}_L}+(1-z){{\bf{u}}_R}}{2},
\quad {\bf{u}}_R^*=\frac{(1+z){{\bf{u}}_R}+(1-z){{\bf{u}}_L}}{2}
\end{equation}
where only the velocity jump at the cell interface are modified by a function $z$. Such a modification plays a role in reducing excess numerical dissipation in low Mach number regions. The function $z$ is determined by the local Mach number,
\begin{equation}\label{eq3.3.2}
z=\min (M_{local},1), \quad M_{local}=\max (M_L,M_R).
\end{equation}	
with
\begin{equation}\label{eq3.3.3}
M_K=\frac{{\sqrt {u_K^2 + v_K^2 + w_K^2} }}{{{a_K}}}, \quad  \text{for} \quad K=L,R .
\end{equation}
It will be demonstrated by numerical results that this low Mach extension method expressed in (\ref{eq3.3.1}) is able to produce physically correct solutions in low Mach number limit.
However, it should be noted that the low Mach number extension method (\ref{eq3.3.1}) will introduce small disturbances in the vicinity of shocks in certain cases \cite{L2Roe}. Thus, the low Mach extension should be turned off around strong shocks. To this end, a further improvement of the low Mach extension method is proposed as
\begin{equation}\label{eq3.3.4}
	{\bf{u}}_K^{\rm{AS}} = f_p \cdot {\bf{u}}_K^ * + (1 - f_p) \cdot {{\bf{u}}_K}, \quad  \text{for} \quad K=L,R .
\end{equation}
where ${\bf{u}}_K^{\rm{AS}}$ represents the velocity used to compute {\bf{A}}ll-{\bf{S}}peed HLL-type schemes defined in (\ref{eq3.1_4}), (\ref{fp}) and (\ref{eq3.3.1})$ \sim $(\ref{eq3.3.4}), $f_p$ denotes the shock detection function defined in (\ref{sd1}) and (\ref{sd2}). In the current study, we call these two schemes ASHLLEM and ASHLLC.

\section{Properties of the modified flux}
\label{S:4}
In the above section, a unified construction framework of all-speed HLL-type schemes is developed by combining a shock stabilization method and a low Mach number modification. The resulting numerical schemes are supposed to compute all-speed flows ranging from low-Mach incompressible to hypersonic flow regimes accurately and robustly. In the following, two important properties of the all-speed HLL-type schemes are examined by both numerical analysis and experiments. First, boundary-layer resolution of the all-speed schemes is discussed, because it is very important for viscous flow computation, especially the heating issue. Then, the shock stability property of the all-speed schemes is explored by a linear perturbation analysis, by which the mechanism of improving the shock robustness by the pressure dissipative flux is discussed in a qualitative way. For the low Mach number performance, a series of numerical experiments is conducted to demonstrate the accuracy of the all-speed schemes. It will be shown in the next section instead of the current one.

\subsection{Contact discontinuity and boundary layer}
\label{S:4.1}
To accurately predict surface heat transfer, it is vital for numerical methods to resolve boundary layer with minimal dissipation, because the heat flux is proportional to temperature gradient in the boundary layer. As we know, the original HLLEM and HLLC solvers are both able to resolve the boundary layer accurately. Thus, we must clarify whether the modification terms proposed in (\ref{eq3.1_4}) and (\ref{eq3.3.4}) will bring negative effects on the resolution of boundary layer flow. To this end, an isolated stationary contact surface is first considered, because such a simple discontinuity is numerically equivalent to a limiting case of a viscous boundary layer, the accuracy of a numerical method for Navier-Stokes equations can be verified by the contact discontinuity problem \cite{park2003dissipation}. There is
\begin{equation}\label{eq4.1.1}
\rho_L \ne \rho_R, \quad q_L=q_R=0, \quad p_L=p_R .
\end{equation}
Considering the relation (\ref{eq4.1.1}), it can be observed from (\ref{sd1}) and (\ref{sd2}) that the pressure-based shock sensing function takes the value of one, thus the pressure dissipative term ${{\bf{F}}_p}$ vanishes. Similarly, the low-Mach modification presented in (\ref{eq3.3.1})$ \sim $(\ref{eq3.3.4}) makes no difference to the original solvers. In conclusion, the unified all-speed construction method introduces no effect on the resolution of contact discontinuity.
\begin{figure}[htbp]
	\centering
	\includegraphics[width=0.70\textwidth]{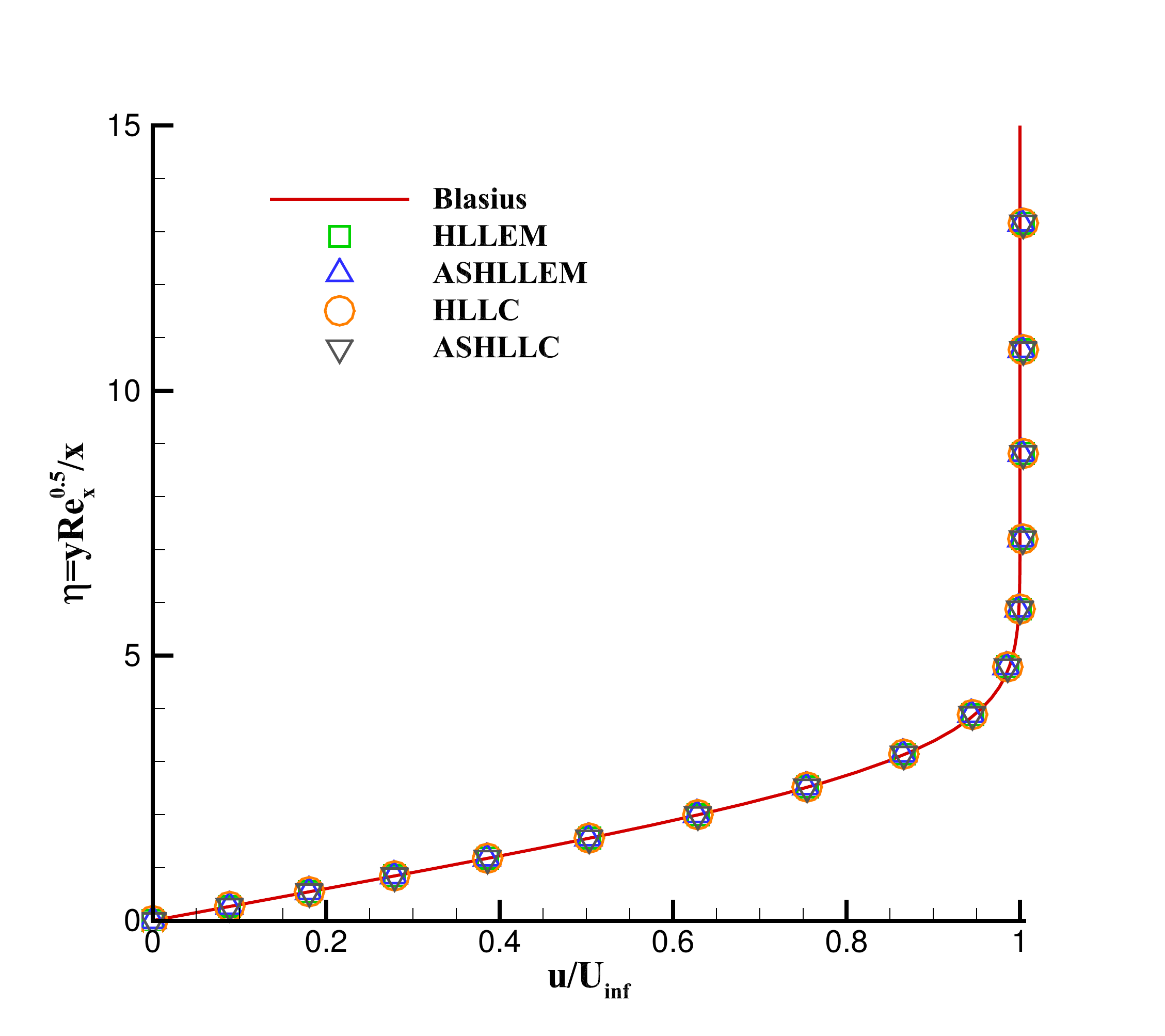}
	\caption{Nondimensional velocity profiles for $Ma=0.3$ laminar boundary layer problem.}
	\label{fig4.1}
\end{figure}
To further assess the capability of all-speed HLL-type schemes in boundary-layer resolving, a laminar boundary layer problem is simulated by a second-order Navier-Stokes code with ASHLLC and ASHLLEM schemes. Readers are referred to \cite{xie2019accurate} for detail descriptions of the numerical setup. Computational results of different schemes are compared in Fig. \ref{fig4.1}. It is shown that the proposed all-speed schemes produce nearly identical results as their unmodified versions.

\begin{figure}
	\centering
	\begin{minipage}[b]{0.70\textwidth}
		\includegraphics[width=1.0\textwidth]{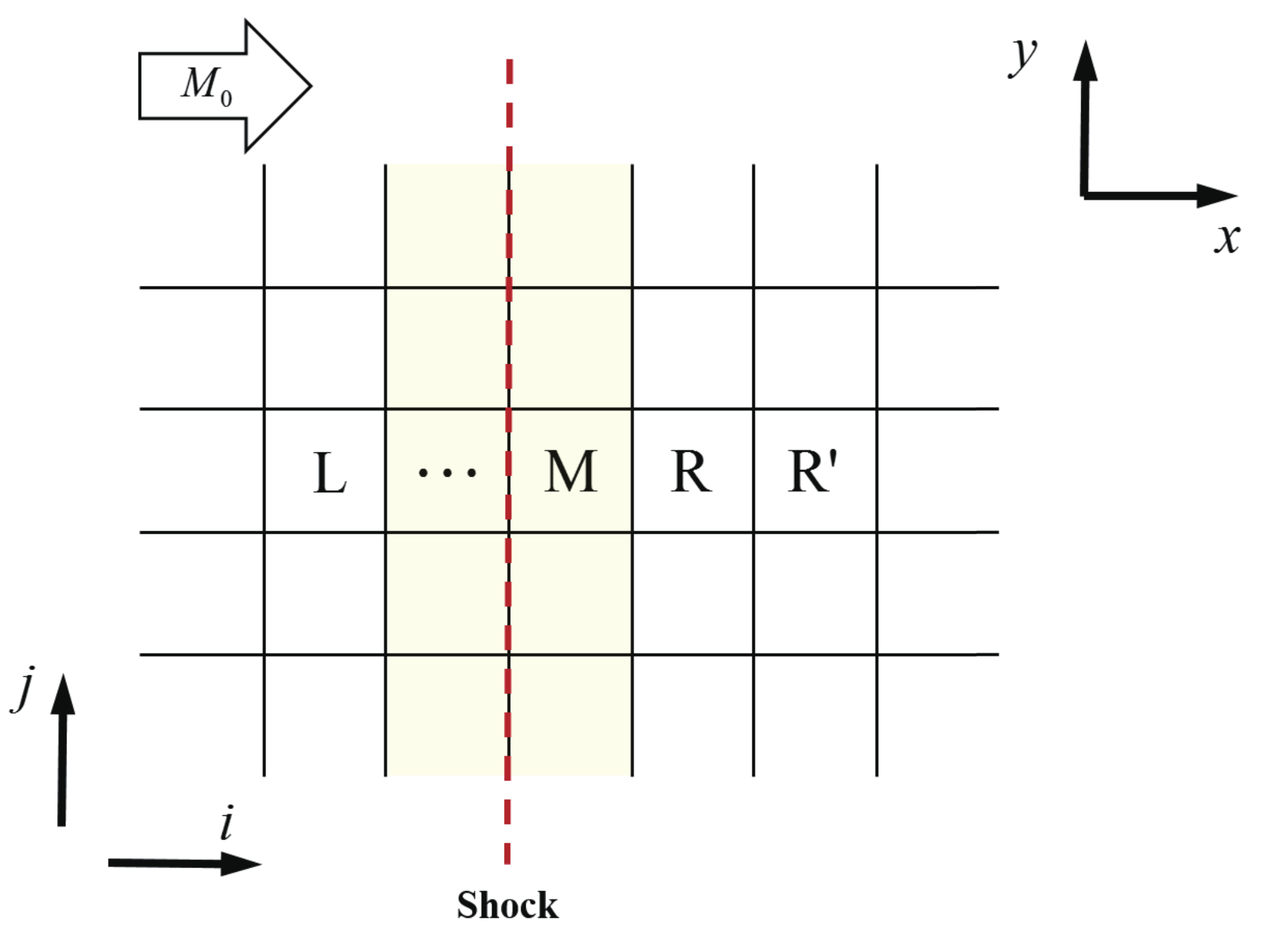}
	\end{minipage}
\caption{Schematic illustration of the numerical shock structure for the steady planar shock problem. L: upstream states, M: intermediate states inside the shock structure, R: downstream states }
\label{fig4.2_1}
\end{figure}

\subsection{Shock stability property : role of pressure dissipative flux}
\label{S:4.2}
So far, we have demonstrated the desirable performance of the pressure dissipative flux and the low-Mach fix for resolving the contact discontinuity and the boundary layer. In this section, we continue to assess the performance of the pressure dissipative flux for shock wave computations and clarify its role for improving shock robustness. To this end, a linearized perturbation analysis is used to analyze the stability property of all-speed HLL-type schemes to solve the planar steady shock. This analysis method is first used by Quirk \cite{quirk1994contribution} and then followed by many other researchers \cite{kim2003cures,Shen2014,xie2017numerical,simonnumerical} to explore the mechanism of numerical shock instabilities. One advantage of the linearized perturbation analysis is that it is able to provide us an intuitive way to understand the mechanism of shock instability in the view of perturbations. In Fig. {\ref{fig4.2_1}}, the structure of a steady planar shock wave in two dimensions is described. As we know, the shock instability is closely related to perturbations and their propagation in the vicinity of strong shocks. Thus, we need to clarify what is the effect of the pressure dissipative flux on the perturbations and their propagation near the shock.

\subsubsection{Propagation of perturbations in shock-normal direction}
\label{sec4.2.1}
In our previous work \cite{xie2017numerical}, it has been found that if the mass flux across the normal shock is accurately preserved (i.e., $(\rho u)^n_L=(\rho u)^n_R$), then the shock could be stabilized. The erroneous mass flux is originated from the intermediate states inside the shock structure. Thus, we need to clarify how the pressure dissipative flux could influence the mass flux perturbation behind the shock. Here, the linearized perturbation analysis is conducted for the numerical flux functions defined in Eq. (\ref{eq3.1_4}). Readers are referred to references \cite{pandolfi2001numerical,xie2017numerical} for the detailed implementation of the analysis.

At time $t^n$, the instability happens. It is assumed that there are some small perturbations being generated in the cell M, they are expressed as
\begin{equation}\label{eq4.2.1.1}
  \rho_M^n=\rho_M^{*,n}+\delta \rho_M^n, \quad (\rho u)_M^n=(\rho u)_M^{*,n}+\delta (\rho u)_M^n, \quad p_M^n=p_M^{*,n}+\delta p_M^n,
\end{equation}
where $\delta (\cdot)$ denote the perturbations which represent small discrepancies from the stable steady states. ${\left(  \cdot  \right)^ *_M }$ denote the stable steady states in cell M, which can be calculated from stable steady states ${\bf{U}}^*_L$ and ${\bf{U}}^*_R$.

To explore how the perturbations generated in cell M influence the mass flux perturbation in cell R, we need to consider the update to momentum component $(\rho u)_R$ in cell R from $x$-direction, that is
\begin{equation}\label{eq4.2.1.2}
  \left( {\rho u} \right)_R^{n + 1} = \left( {\rho u} \right)_R^n - \frac{{\Delta t}}{{\Delta x}}\left[ {\left( {\rho {u^2} + p} \right)_{R,R'}^n - \left( {\rho {u^2} + p} \right)_{M,R}^n} \right].
\end{equation}
The subscript $R,R'$ denotes the interface between the cell $R$  and the cell $R'$ , the subscript $M,R$  denotes the interface between the cell $M$  and the cell $R$. Here, the numerical flux functions defined in Eq. (\ref{eq3.1_4}) are used to solve the momentum fluxes at the interfaces in Eq. (\ref{eq4.2.1.2}). For the original flux functions HLLEM and HLLC defined in (\ref{eq2.2.1_7}) and (\ref{2.7}), the update to the perturbed momentum component $\delta \left( {\rho u} \right)_R$ can be written as the following form,
\begin{equation}\label{eq4.2.1.3}
\delta \left( {\rho u} \right)_R^{n + 1} - \delta \left( {\rho u} \right)_R^n = {\theta}_{\rho} \cdot \delta {\rho}_M^n + {\theta}_{u} \cdot \delta u_M^n + {\theta}_p \cdot \delta p_M^n
\end{equation}
where the coefficients ${\theta}_{\rho}$, ${\theta}_u$ and ${\theta}_p$ are functions of the freestream Mach number $M_0$ and the conservative variables ${\bf{U}}^*_L$ and ${\bf{U}}^*_R$. Here, the high order small perturbations ${\delta}^k(\cdot)_{k\ge2}$ are negligible and omitted during the calculation. These coefficients differ by numerical flux functions to solve the momentum components at interfaces. Here, the exact formulas of these coefficients are omitted, because we only need concern the update to the perturbed momentum component $\delta \left( {\rho u} \right)_R$ from the pressure dissipative flux ${\bf{F}}_p$. Inserting Eq. (\ref{eq4.2.1.1}) and Eq. (\ref{fp}) into Eq. (\ref{eq4.2.1.2}), the resulting evolution for the perturbed mass flux in $x$-direction can be written as,
\begin{equation}\label{eq4.2.1.4}
  \delta \left( {\rho u} \right)_R^{n + 1} - \delta \left( {\rho u} \right)_R^n = {\xi _p} \cdot \delta p_M^n,
\end{equation}
with
\begin{equation}\label{eq4.2.1.5}
  {\xi _p} = \left( {1 - {f_p}} \right)\frac{{{S_M}{S_R}}}{{{S_R} - {S_M}}}\frac{{\nu {u^{*,n}}}}{{{a^{*,n}}{{\left( {{u^{*,n}} + {a^{*,n}}} \right)}^2}}} ,
\end{equation}
where $\nu$ denotes the Courant number, $S_M$ and $S_R$ denote the left and right wavespeeds evaluated by ${\bf{U}}^*_M$ and ${\bf{U}}^*_R$ respectively and they are assumed to be unperturbed during the calculation. ${u^{*,n}}$ and ${a^{*,n}}$ represent stable steady velocity and sound speed.

The update to the total perturbed momentum component ${\delta(\rho u)}_R$ from $x$-direction can be obtained by combining Eq. (\ref{eq4.2.1.3}) and Eq. (\ref{eq4.2.1.4}),
\begin{equation}\label{eq4.2.1.6}
\delta \left( {\rho u} \right)_R^{n + 1} - \delta \left( {\rho u} \right)_R^n = {\theta}_{\rho} \cdot \delta {\rho}_M^n + {\theta}_{u} \cdot \delta u_M^n + ({\theta}_p + {\xi _p}) \cdot \delta p_M^n .
\end{equation}
In the vicinity of shocks, $f_p$ is always smaller than unit and the wave speed $S_M$ remains nonpositive, thus it can be observed from (\ref{eq4.2.1.5}) that the coefficient $\xi _p$ remains nonpositive. Considering the relation in (\ref{eq4.2.1.6}), it can be obtained that the pressure dissipative flux plays a role in reducing the pressure perturbation that contributes to the erroneous mass flux. During the calculation of the steady planar shock, the upstream states in cell L remain unperturbed, thus a reduced erroneous mass flux in cell R is helpful for maintaining the mass flux consistence, i.e., $(\rho u)^n_L=(\rho u)^n_R$. As a result, the shock instability can be suppressed.

\subsubsection{Propagation of perturbations in shock-tangential direction}
\label{sec4.2.2}
The magnitude of the velocity in the transverse direction of the shock wave has been well recognized to be a proper parameter to use to show the magnitude of the multidimensional carbuncle phenomenon \cite{Dumbser2004,henderson2007grid}. For the steady normal shock problem, physically, there should be no mass flux appearing in the transverse direction and the transverse velocity should be zero. Hence, any erroneous mass flux developed in this direction results from the instability.

At the beginning of the instability, perturbations are generated inside the shock structure. To facilitate further analysis, it is assumed that the states along the $y$ direction inside the shock structure are initialized as follows,
\begin{equation}\label{eq4.2.2.1}
\rho_{j}^n=\rho_{i,j}^{*,n}-\delta {\rho}^n, \quad \left( {\rho u} \right)_{i,j}^n = {\left( {\rho u} \right)_{i,j}^{*,n}}-\delta ({\rho u})^n, \quad \left( {\rho v} \right)_{i,j}^n = {\left( {\rho v} \right)_{i,j}^{*,n}} - \delta ({\rho v})^n, \quad p_{i,j}^n = {p_{i,j}^{*,n}} - \delta p^n,
\end{equation}
and
\begin{equation}\label{eq4.2.2.2}
  \rho _{i,j \pm 1}^n = \rho_{i,j \pm 1}^{*,n} + \delta {\rho}^n, \quad \left( {\rho u} \right)_{i,j \pm 1}^n = {\left( {\rho u} \right)_{i,j \pm 1}^{*,n}}+\delta ({\rho u})^n, \quad \left( {\rho v} \right)_{i,j \pm 1}^n = {\left( {\rho v} \right)_{i,j \pm 1}^{*,n}} + \delta ({\rho v})^n, \quad p_{i,j \pm 1}^n = {p_{i,j \pm 1}^{*,n}} + \delta {p^n},
\end{equation}
where ${()}^{*}$ represent the stable steady solutions that are assumed to be uniform along the transverse direction. In what follows, we omit the subscript $i$ for clarity. In the two-dimensional case, we need to clarify how the perturbations will promote the perturbed mass flux in the transverse direction. Hence, the following conservative scheme is considered,
\begin{equation}\label{eq4.2.2.3}
  \left( {\rho v} \right)_j^{n + 1} = \left( {\rho v} \right)_j^n - \frac{{\Delta t}}{{\Delta y}}\left[ {\left( {\rho {v^2} + p} \right)_{j + 1/2}^n - \left( {\rho {v^2} + p} \right)_{j - 1/2}^n} \right].
\end{equation}
Similarly, the improved flux functions defined in (\ref{eq3.1_4}) are used to solve the momentum flux at the interfaces in (\ref{eq4.2.2.3}). For the original HLLEM and HLLC flux functions, the update to the perturbed momentum component can also be rewritten as the following form,
\begin{equation}\label{eq4.2.2.4}
    \delta \left( {\rho v} \right)_j^{n + 1} - \delta \left( {\rho v} \right)_j^{n} = {\theta _\rho } \cdot \delta \rho ^n{\rm{ + }}{\theta _v} \cdot \delta v^n + {\theta _p} \cdot \delta p^n
\end{equation}
where the coefficients ${\theta}_{\rho}$, ${\theta}_v$ and ${\theta}_p$ differ by numerical flux functions to solve the momentum components at interfaces. The exact formulas of these coefficients are still not presented and we only concern the update to the perturbed mass flux $\delta \left( {\rho v} \right)_j$ due to the pressure dissipative flux ${\bf{F}}_p$, that is
\begin{equation}\label{eq4.2.2.5}
  \delta \left( {\rho v} \right)_j^{n + 1} - \delta \left( {\rho v} \right)_j^n = {\xi _p} \cdot \delta {p^n} ,
\end{equation}
with
\begin{equation}\label{eq4.2.2.6}
  {\xi _p} = 4\left( {1 - {f_p}} \right)\frac{{{S_L}{S_R}}}{{{S_R} - {S_L}}}\frac{{\nu {v^{*,n}}}}{{{a^{*,n}}{{\left( {{v^{*,n}} + {a^{*,n}}} \right)}^2}}} ,
\end{equation}
where $\nu$ denotes the Courant number, $S_L$ and $S_R$ denote the left and right wavespeeds evaluated by ${\bf{U}}^*_L$ and ${\bf{U}}^*_R$ respectively and they are assumed to be unperturbed during the calculation. ${v^{*,n}}$ and ${a^{*,n}}$ represent stable steady velocity and sound speed. The update to the total perturbed momentum component ${\delta(\rho v)}_j$ from $y$-direction can be obtained by combining Eq. (\ref{eq4.2.2.4}) and Eq. (\ref{eq4.2.2.5}),
\begin{equation}\label{eq4.2.2.7}
\delta \left( {\rho v} \right)_j^{n + 1} - \delta \left( {\rho v} \right)_j^n = {\theta}_{\rho} \cdot \delta {\rho}^n + {\theta}_{v} \cdot \delta v^n + ({\theta}_p + {\xi _p}) \cdot \delta p^n .
\end{equation}
In the vicinity of shocks, $f_p$ is always smaller than unit and the wave speed $S_L$ remains nonpositive, thus it can be observed from (\ref{eq4.2.2.6}) that the coefficient $\xi _p$ remains nonpositive. Considering the relation in (\ref{eq4.2.2.7}), it can be obtained that the pressure dissipative flux plays a role in reducing the pressure perturbation that contributes to the erroneous mass flux in the $y$ direction. As a result, the shock instability can be suppressed.

\section{Numerical results}
\label{S:5}
In this section, the proposed all-speed HLL-type schemes are applied to a series of numerical experiments, a few of which have already been used for the validation of several improved Godunov-type schemes \cite{xie2019accurate,xie2019towards}. The flow regimes range from low Mach incompressible flows to hypersonic compressible flows. A major focus is on assessing the accuracy and robustness of the proposed all-speed schemes for hypersonic heating problem.

\subsection{Low Mach number flows around the NACA 0012 airfoil}
\label{S:5.1}
To purely examine the performance of the proposed all-speed HLL-type schemes at a low Mach number flow, we consider the classical inviscid flow about the NACA 0012 airfoil. For the normalized Euler equations, it is well known that the discrete solutions of first-order HLL-type schemes support pressure fluctuations in space of order $M_{0}$, with $p\left( {{\bf{x}},t} \right) = {p_0}\left( t \right) + {p_1}\left( {{\bf{x}},t} \right){M_0}$. Whereas the continuous pressure fluctuations scale as $M_{0}^2$. As a result, without proper rescaling of the numerical dissipation in HLL-type schemes, a compressible flow solver usually fails to provide accurate results for low Mach number flows.
\begin{figure}
	\centering
	\subfigure[HLLEM]{
	\begin{minipage}[b]{0.48\textwidth}
		\includegraphics[width=1.0\textwidth]{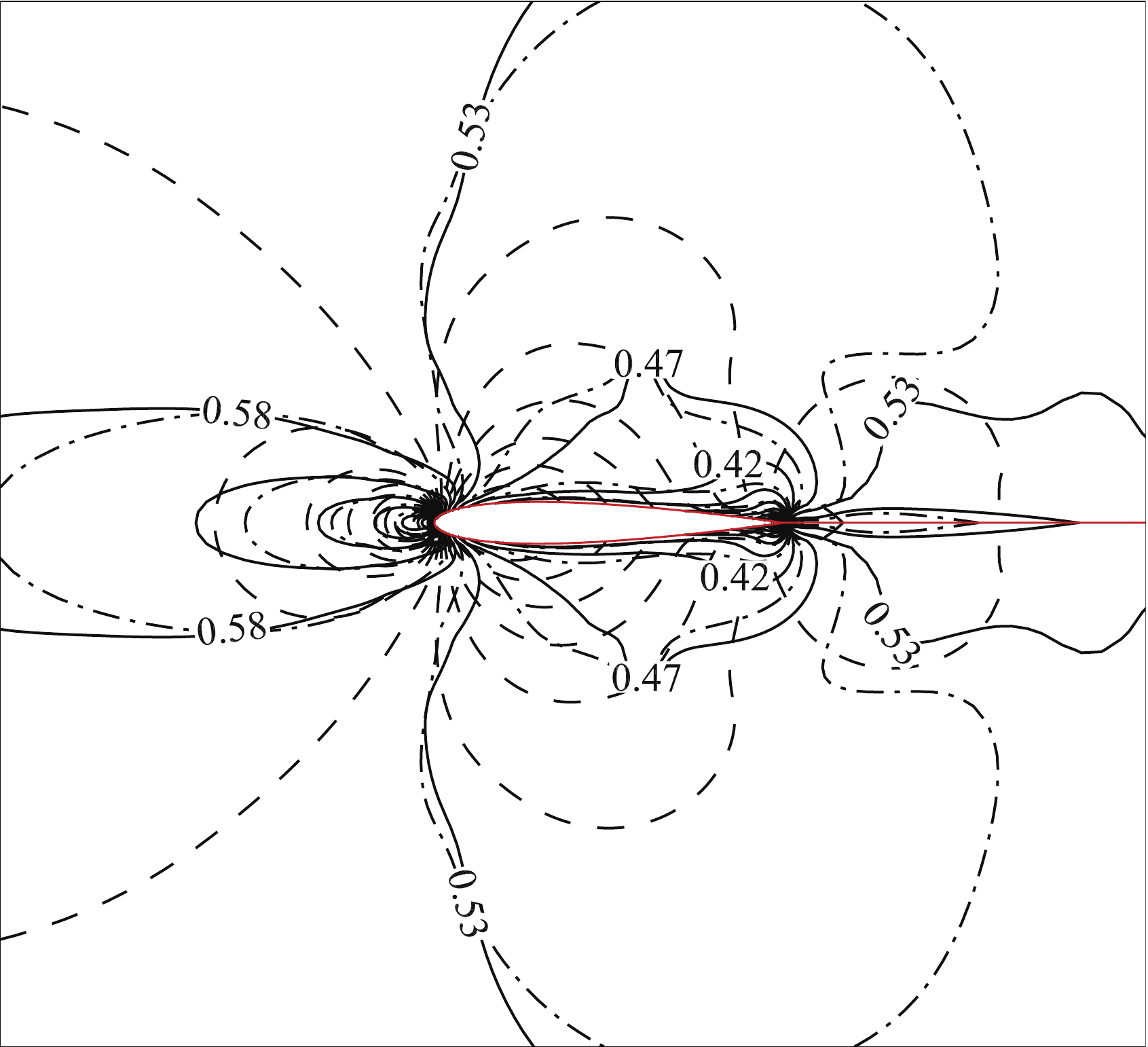}
	\end{minipage}
	}
	\subfigure[ASHLLEM]{
	\begin{minipage}[b]{0.48\textwidth}
		\includegraphics[width=1.0\textwidth]{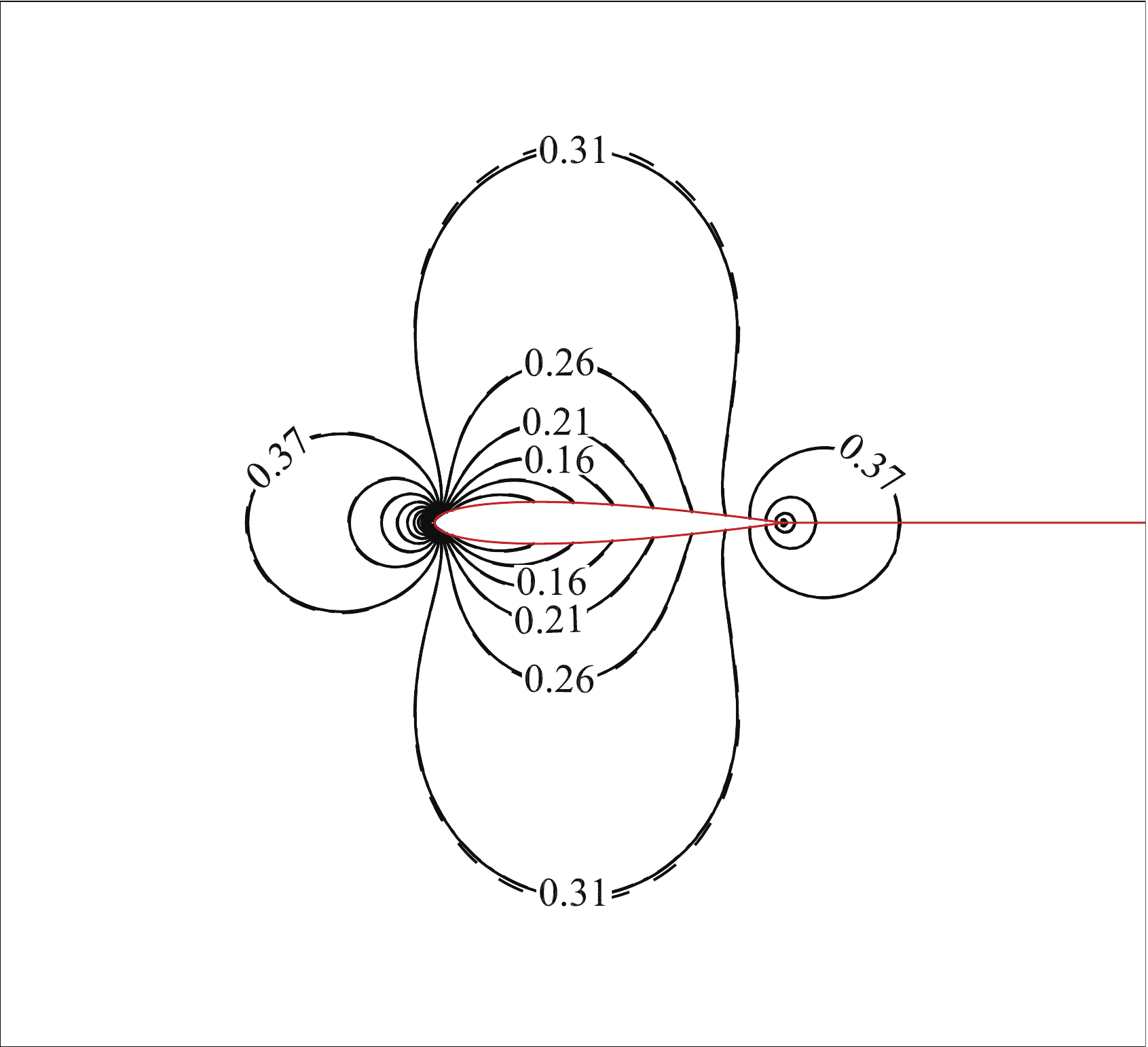}
	\end{minipage}
	}
	\centering
	\subfigure[HLLC]{
	\begin{minipage}[b]{0.48\textwidth}
	\includegraphics[width=1.0\textwidth]{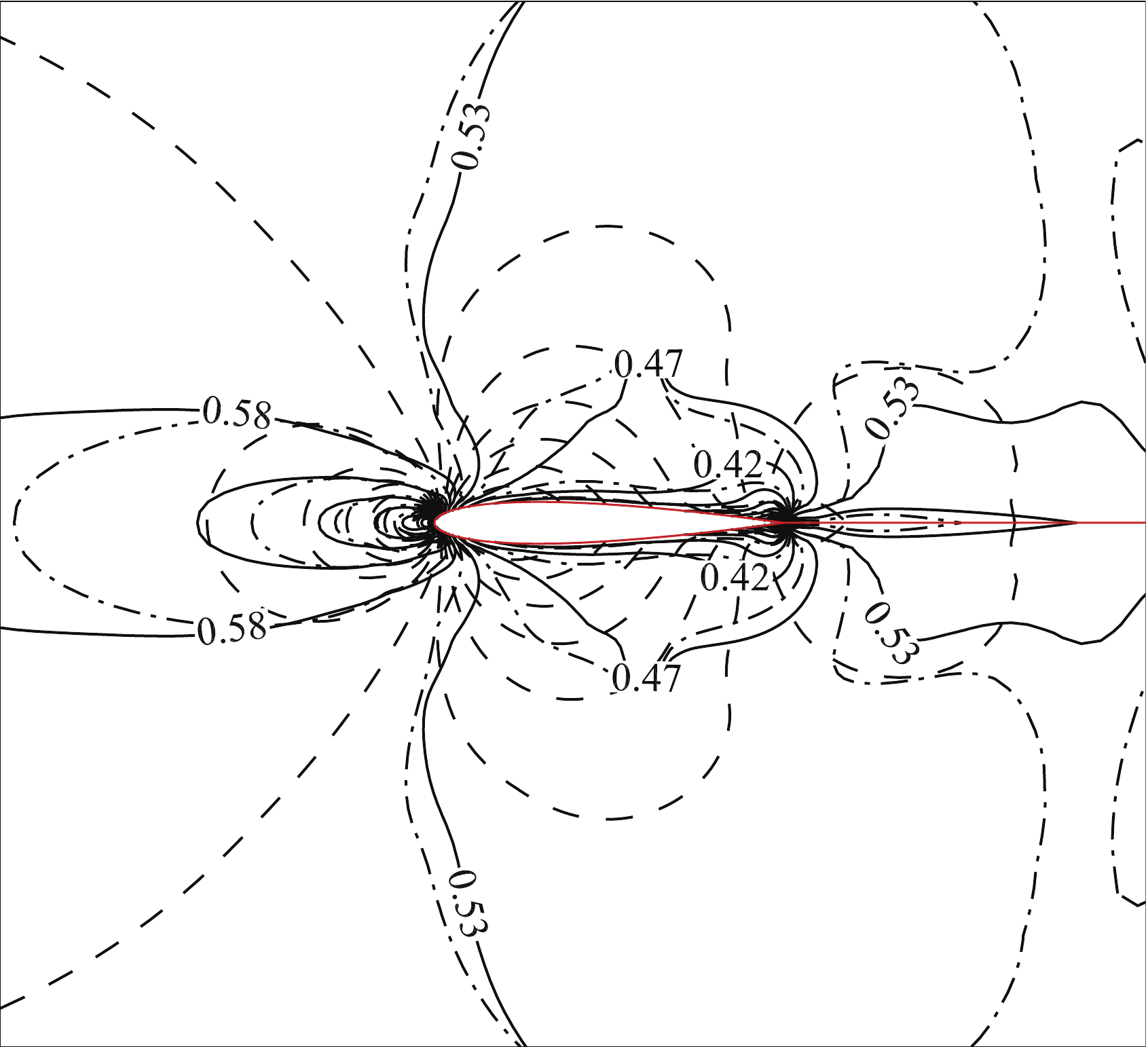}
	\end{minipage}
	}
	\subfigure[ASHLLC]{
	\begin{minipage}[b]{0.48\textwidth}
	\includegraphics[width=1.0\textwidth]{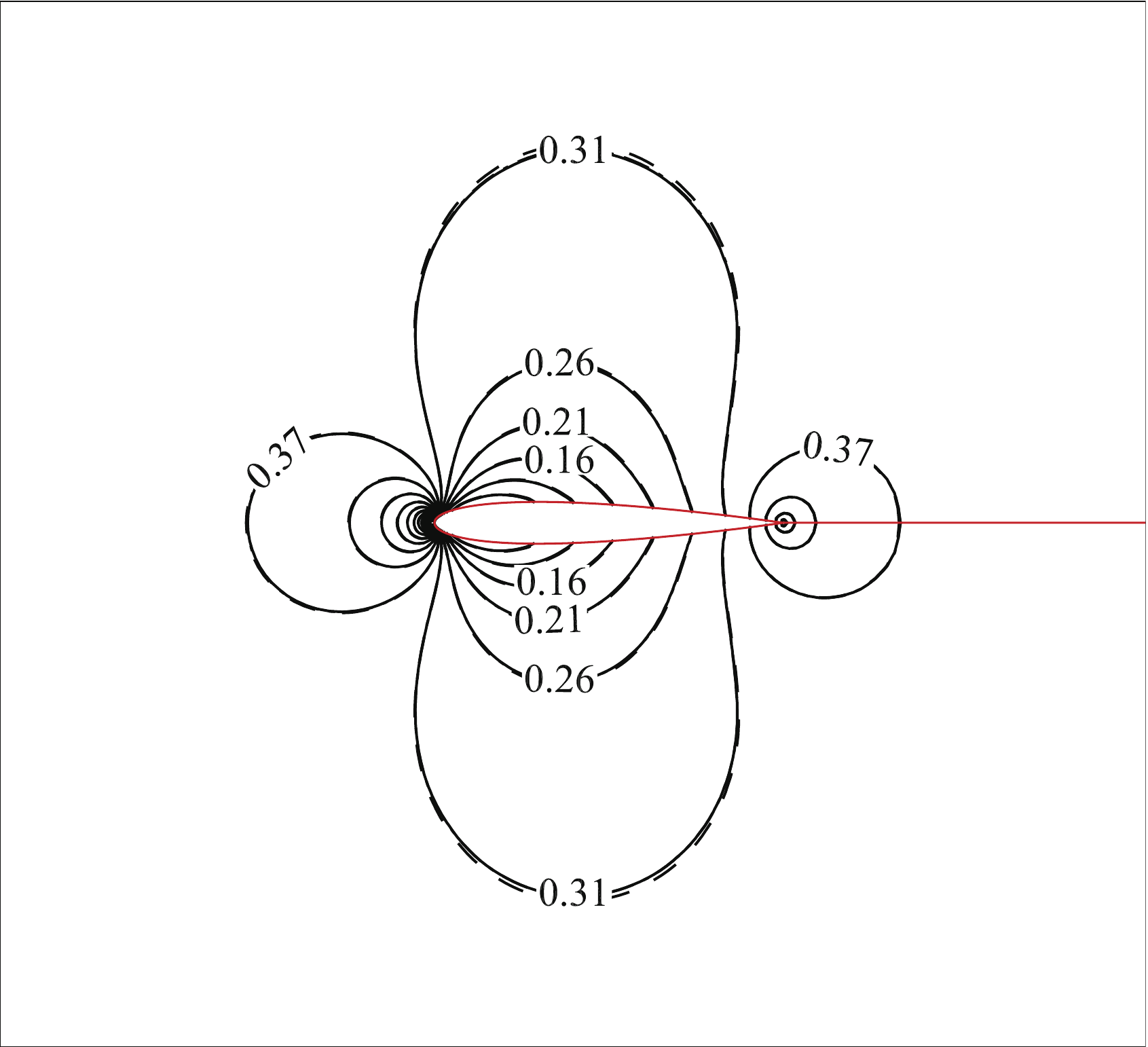}
	\end{minipage}
	}
\caption{Contours of the normalized pressure with the inflow Mach number.
\protect\tikz[baseline]{\protect\draw[line width=0.2mm,loosely dashed] (0,.8ex)--++(0.8,0) ;}~$M_0=10^{-1}$,
\protect\tikz[baseline]{\protect\draw[line width=0.2mm,dash dot] (0,.8ex)--++(0.8,0) ;}~$M_0=10^{-2}$,
\protect\tikz[baseline]{\protect\draw[line width=0.2mm] (0,.8ex)--++(0.8,0) ;}~$M_0=10^{-3}$.
}
\label{fig5.1_1}
\end{figure}

To demonstrate whether the all-speed HLL-type schemes can compute low Mach number flows or not, a sequence of computations with decreasing inflow Mach numbers is carried out on the same O-type mesh. The computational domain is discretized by quadrilateral grids with $241$ (circumferential) $\times$ $121$ (normal) cell number. All the simulations are conducted at zero angle-of-attack with three Mach numbers, $M_0=0.1,\,0.01,\,0.001$, where the steady solutions are obtained after $50,000$ time steps with $\rm{CFL}=100$ using LU-SGS approach. The residuals (L2-norm of density) dropped at least ten orders of magnitude for all the cases.
\begin{figure}
	\centering
	\begin{minipage}[b]{0.70\textwidth}
		\includegraphics[width=1.0\textwidth]{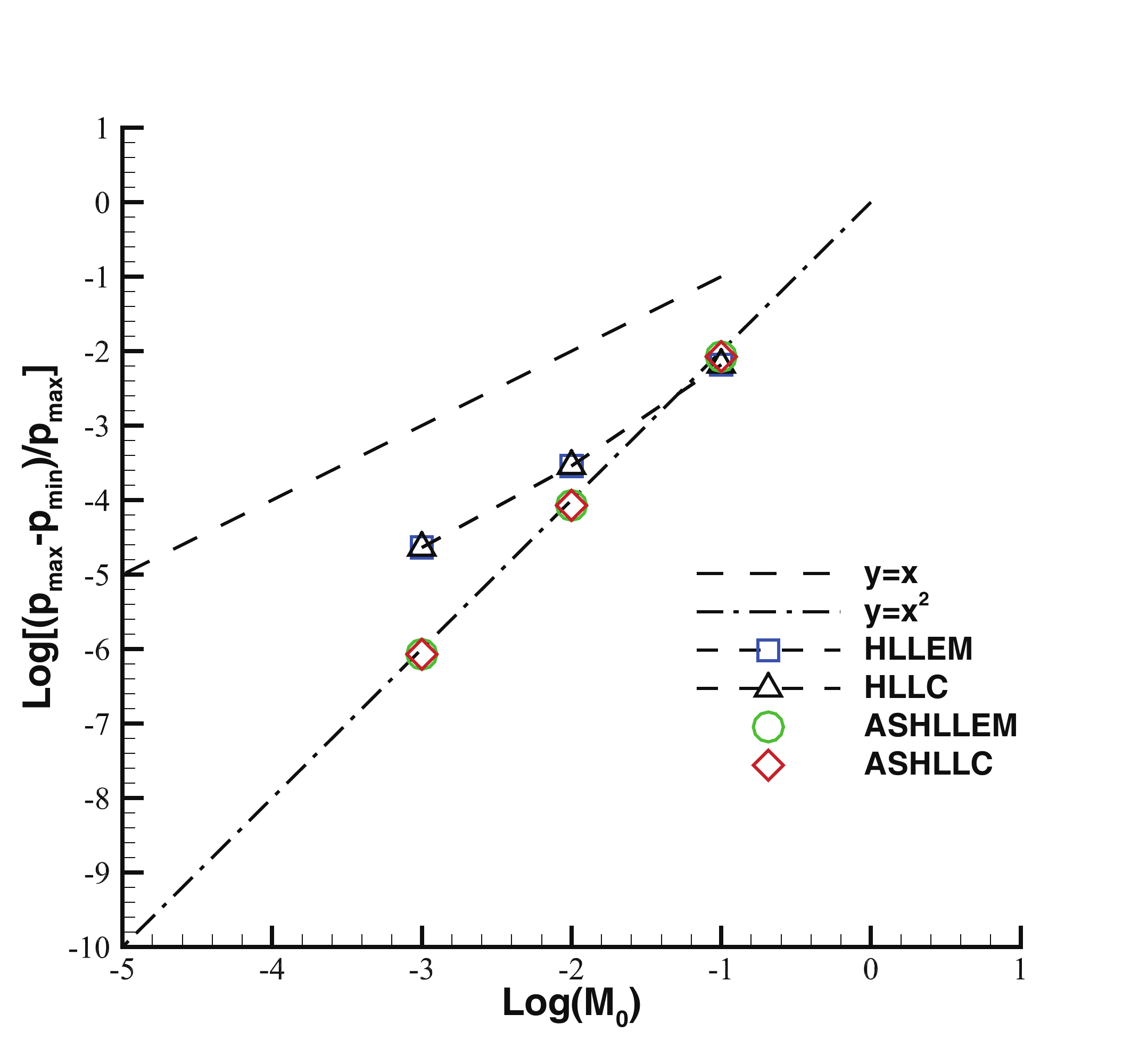}
	\end{minipage}
\caption{Pressure fluctuations with the inflow Mach number for different schemes.}
\label{fig5.1_2}
\end{figure}
Computational results are shown in Fig. \ref{fig5.1_1}, where the normalized pressure \cite{guillard1999behaviour} defined as ${p_N}\left( {\bf{x}} \right) = \frac{{p\left( {\bf{x}} \right) - {p_{\min }}}}{{{p_{\max }} - {p_{\min }}}}$ is used for the comparison of the pressure fields at different inflow Mach numbers. One should notice that the normalized pressure is independent from the inflow Mach number. Thus, as the inflow Mach number approaches zero, numerical solutions should also converge to a consistent approximation of the incompressible solution \cite{boniface2017rescaling}. As shown in Fig. \ref{fig5.1_1}, solutions obtained with ASHLLEM and ASHLLC schemes converge to a unique isentropic solution, which cannot be achieved with the HLLEM and HLLC schemes. The behaviour of the pressure fluctuations with the inflow Mach number is demonstrated in Fig. \ref{fig5.1_2}. As expected, first-order HLLEM and HLLC schemes for compressible flows support fluctuations of order $M_{0}$ in the incompressible limit, whereas the physical pressure fluctuation should scale as $M_0^2$. With the all-speed HLL-type schemes, the pressure fluctuations exactly scale with $M_{0}^2$.

\subsection{RAE 2822 transonic airfoil}
\label{S:5.2}
The second test case considered here is the viscous turbulent flows over RAE 2822 airfoil at the transonic regime \cite{Cook1979}. It is used to demonstrate the accuracy of the all-speed schemes for transonic flow computations and assess the smooth transition of the all-speed HLL-type schemes with their original versions at the sonic line. The computational mesh is a structured O-type grid with dimensions of $369$ (circumferential) $\times$ $165$ (normal) and the minimal mesh size near the airfoil surface is $1\times10^{-5}C$. The outer boundary is at $200$C with the farfield boundary conditions. The freestream Mach number is set as $0.729$ at an angle of attack of $2.31^{\circ}$, and the freestream static temperature is $460.0$R. These conditions correspond to a Reynolds number of $6.5$ million based on the chord length. The static pressure can be computed based on the specified Reynolds number, Mach number and the static temperature. Numerical experiments are conducted with the finite volume method to solve the Reynolds-averaged Navier-Stokes equations with SA turbulence model. All-speed HLL-type schemes and their original versions are used to discrete the convective flux and 2nd MUSCL reconstruction with minmod limiter is applied. All the computations are conducted using the LU-SGS approach with CFL=5 for 100,000 time iterations. A convergence criterion of ten orders of magnitude reduction of the mean-flow equations residual is used.

\begin{figure}
	\centering
	\subfigure[HLLEM and ASHLLEM]{
	\begin{minipage}[b]{0.48\textwidth}
		\includegraphics[width=1.0\textwidth]{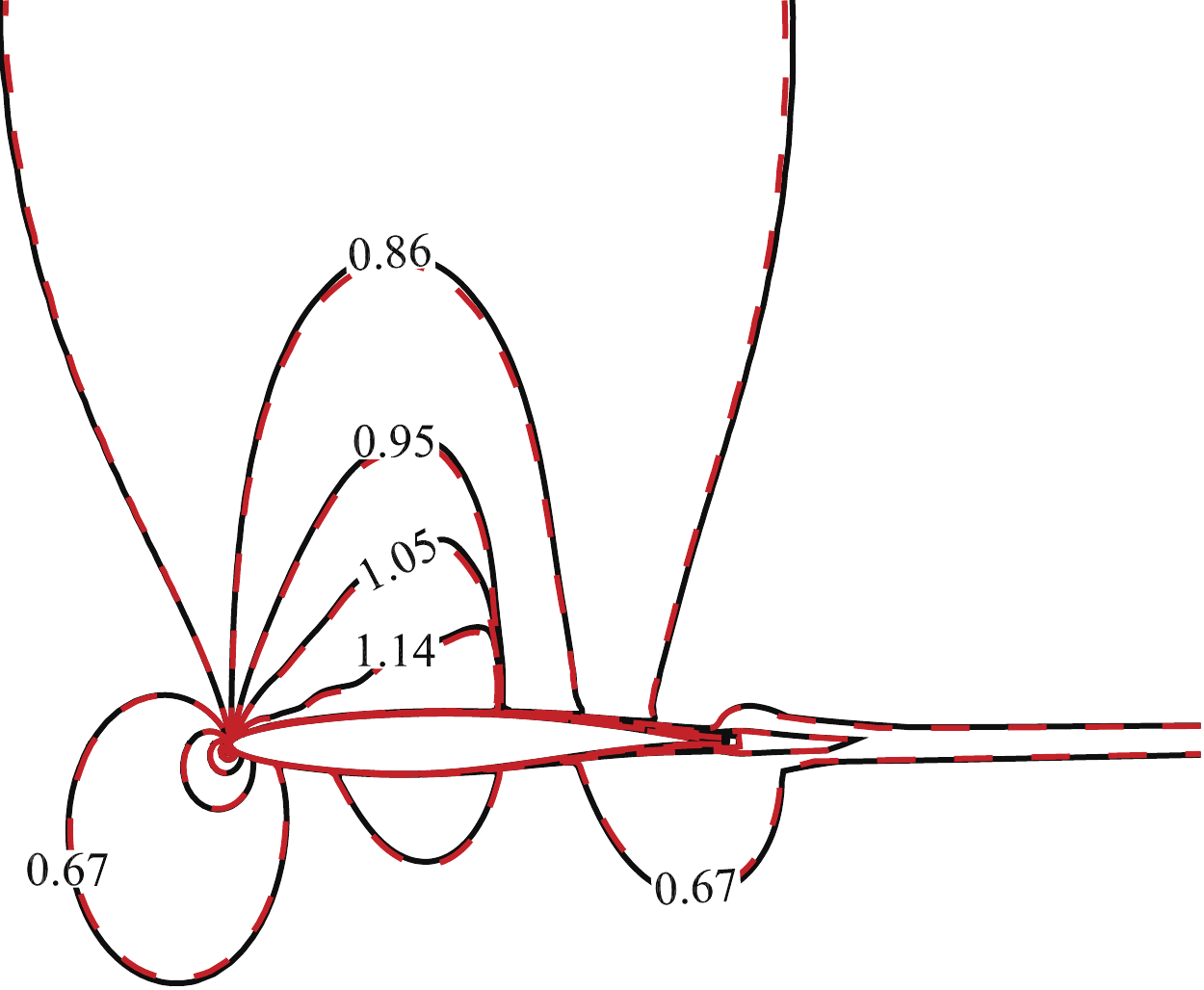}
	\end{minipage}
	}
	\subfigure[HLLC and ASHLLC]{
	\begin{minipage}[b]{0.48\textwidth}
		\includegraphics[width=1.0\textwidth]{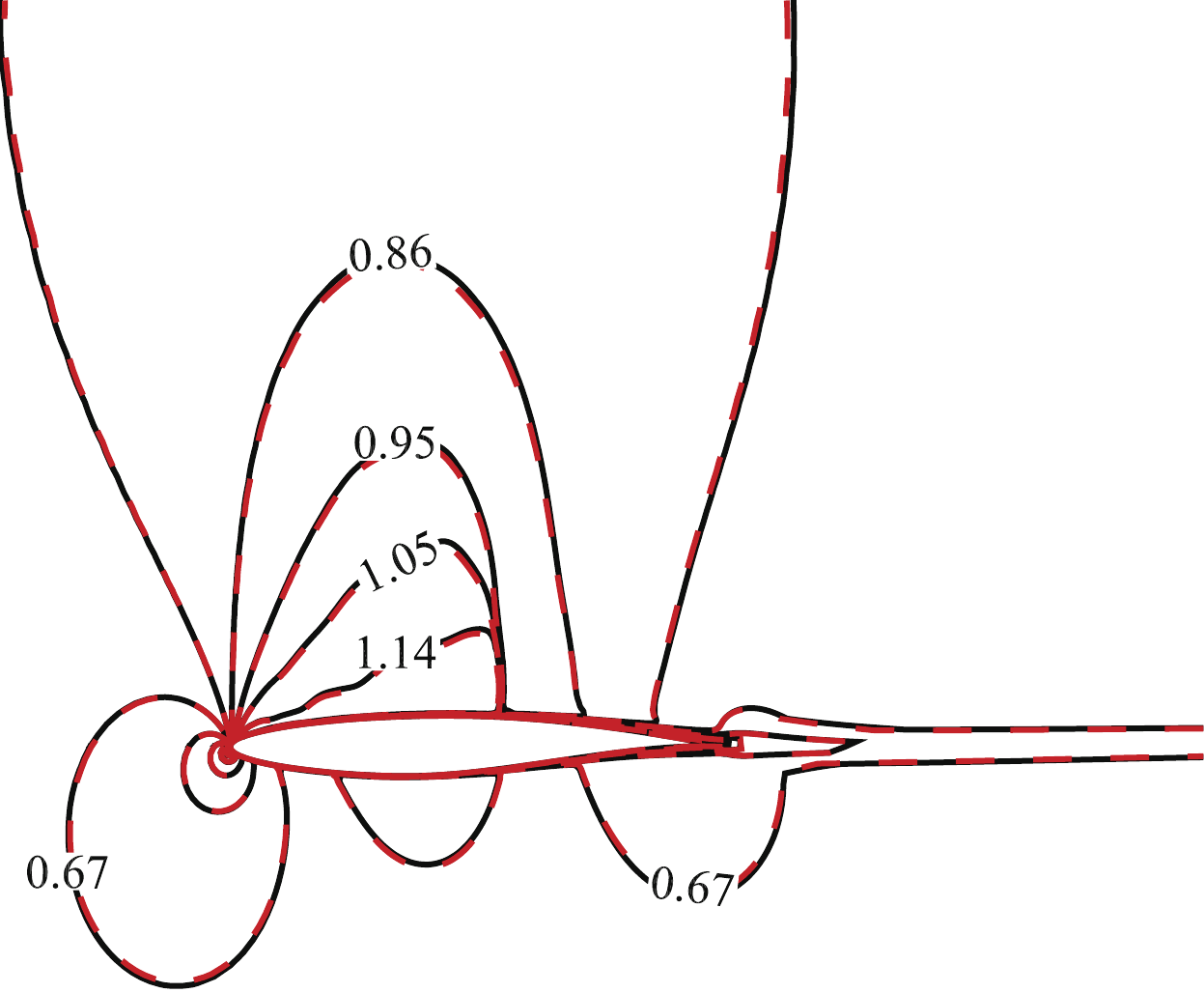}
	\end{minipage}
	}
\caption{Mach number isolines for RAE 2822 airfoil (12 equally spaced levels from 0.1 to 1.2). Original schemes (black solid line) and all-speed schemes (red dashed).}
\label{fig_rae_a}
\end{figure}

\begin{figure}[htbp]
	\centering
	\includegraphics[width=0.75\textwidth]{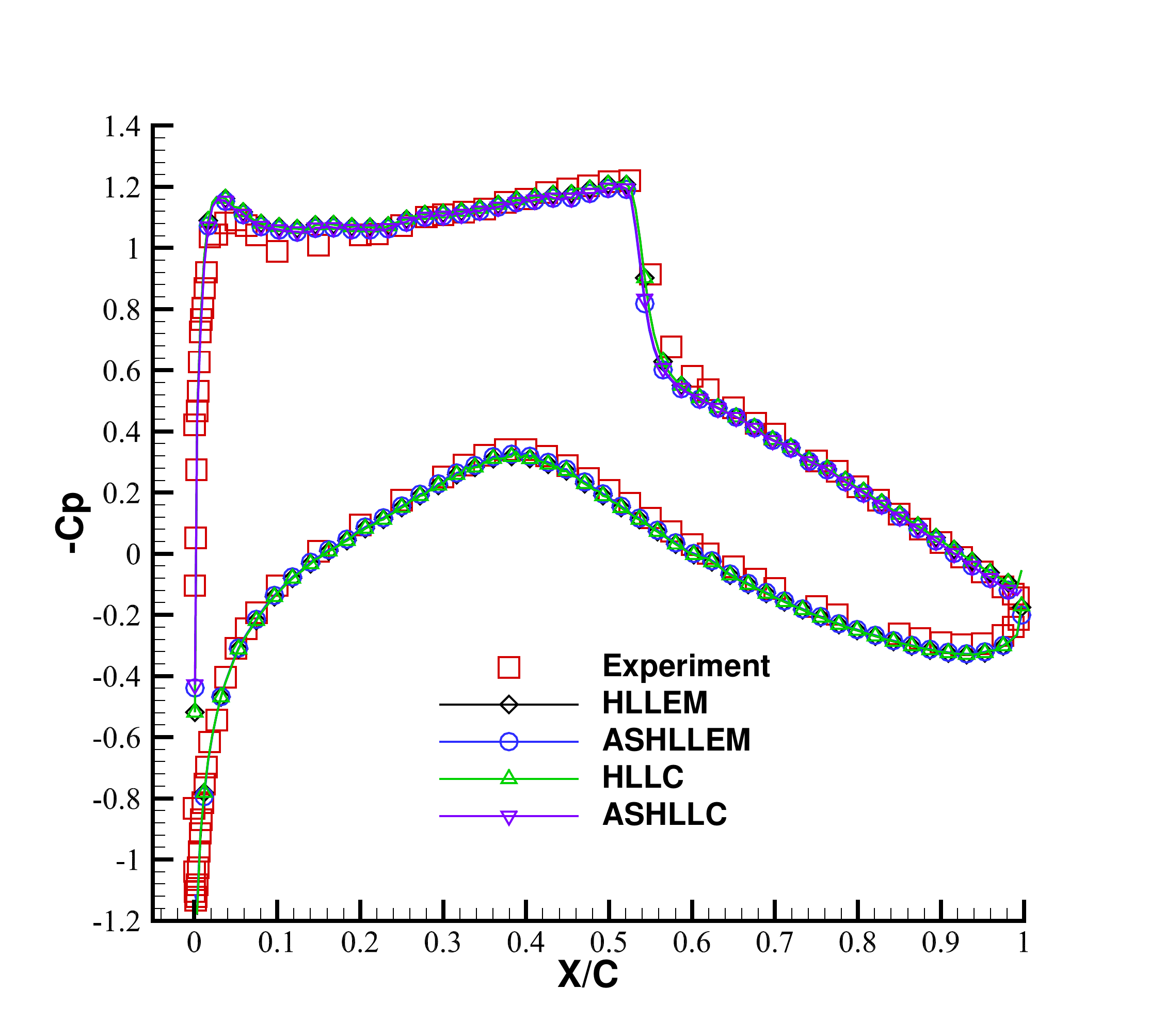}
	\caption{Surface pressure coefficient $C_p$ for RAE 2822 airfoil.}
	\label{fig_rae_b}
\end{figure}

In Fig. \ref{fig_rae_a}, the Mach number isolines are shown. It can be observed that the superimposed solutions of the all-speed HLL-type schemes and their original versions are hardly distinguishable. Surface pressure coefficients are presented in Fig. \ref{fig_rae_b}, where the experimental data are also used for comparison. As shown, both the all-speed schemes resolve the smooth surface pressure coefficient profiles well and there is no any significant difference between the numerical results. They are in good agreement with the experimental data.

\subsection{Double Mach reflection problem}
\label{S:5.3}
In the above two test cases, we have assessed the accuracy and robustness of the all-speed HLL-type schemes in incompressible and transonic flow regimes. In the current section and the following one, we focus on verifying the robustness of all-speed schemes for unsteady and steady hypersonic flows. Here, the double Mach reflection problem is used to demonstrate the shock robustness of all-speed schemes for unsteady hypersonic flow computations. This problem describes a planar shock wave propagating in inviscid fluid which is reflected by a $30^{\circ}$ ramp. It is first studied by Woodward and Colella \cite{Woodward1984} and followed by many other scholars to test numerical behaviours of shock-capturing methods. The computational domain is $\left[0, 4\right] \times \left[0, 1\right]$, which has been divided into 960 cells along the length and 240 cells along the width. The shock wave has a strength with Mach number 10, which is initially set up to be inclined at an angle of 60 with the bottom reflecting wall. The domain in front of the shock is initialized with pre shock values given as $\rho = 1.4$, $u=0$, $v=0$, $p=1$. The domain behind the shock is initialized to post shock values. At the top boundary, the flow variables are set to describe the exact motion of the shock front along the wall. The inflow and outflow boundary conditions are used at the entrance and the exit.

\begin{figure}[htbp]
	\centering
	\subfigure[$\rm{HLLEM}$]{
		\begin{minipage}[b]{0.75\textwidth}
			\centering
			\includegraphics[width=0.8\textwidth]{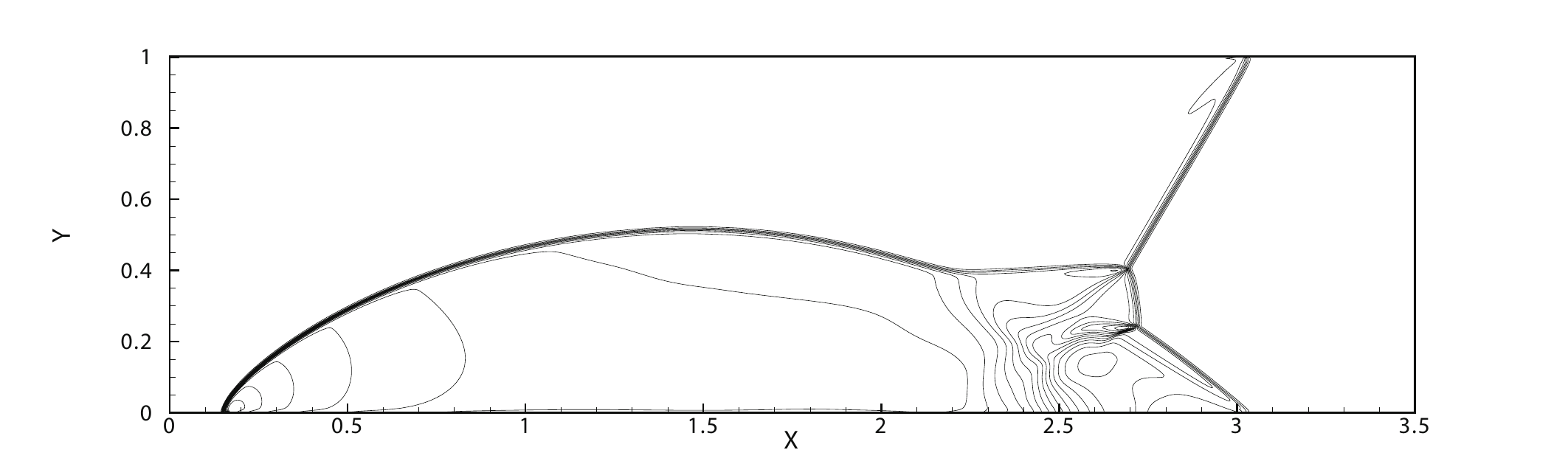}
		\end{minipage}
	}
	\subfigure[$\rm{HLLC}$]{
		\begin{minipage}[b]{0.75\textwidth}
			\centering
			\includegraphics[width=0.8\textwidth]{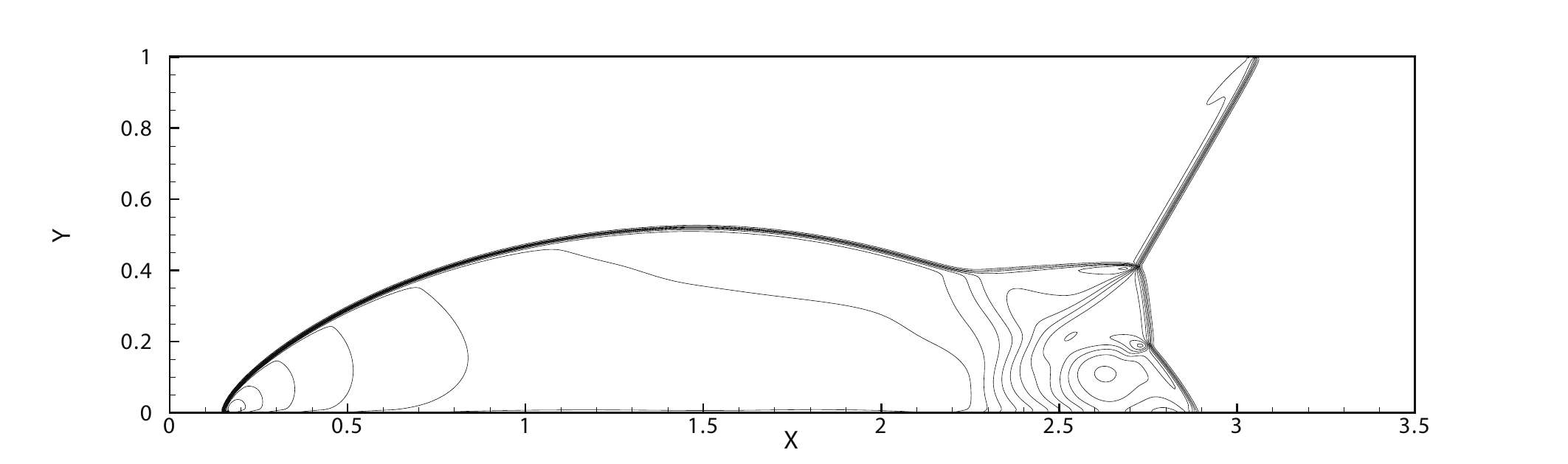}
		\end{minipage}
	}
	\subfigure[$\rm{ASHLLEM}$]{
		\begin{minipage}[b]{0.75\textwidth}
			\centering
			\includegraphics[width=0.8\textwidth]{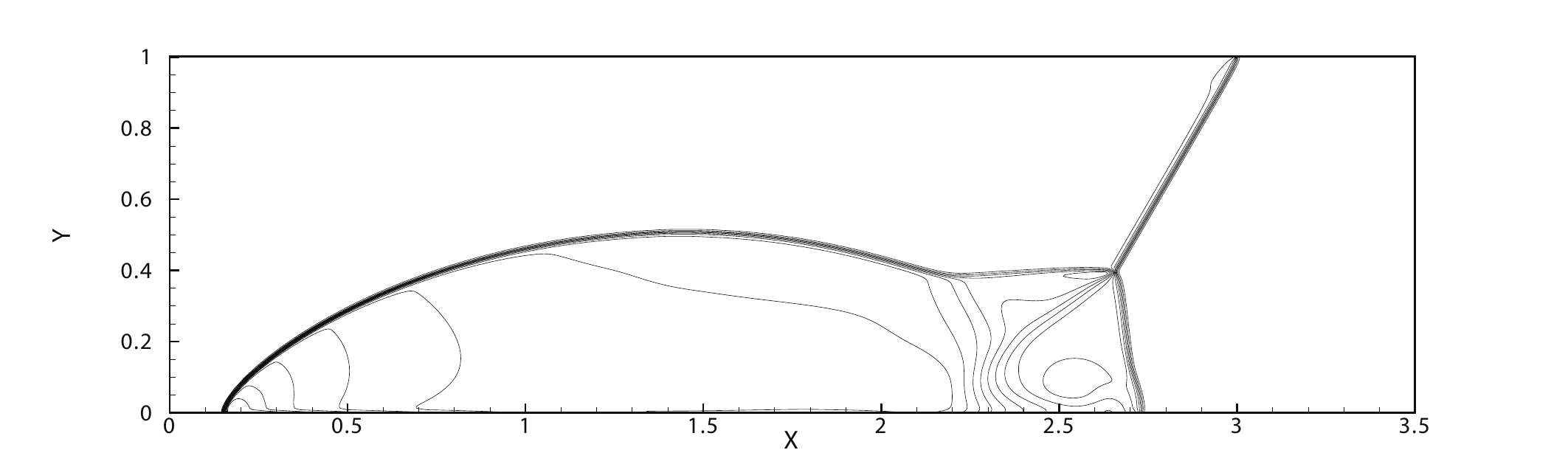}
		\end{minipage}
	}
	\subfigure[$\rm{ASHLLC}$]{
		\begin{minipage}[b]{0.75\textwidth}
			\centering
			\includegraphics[width=0.8\textwidth]{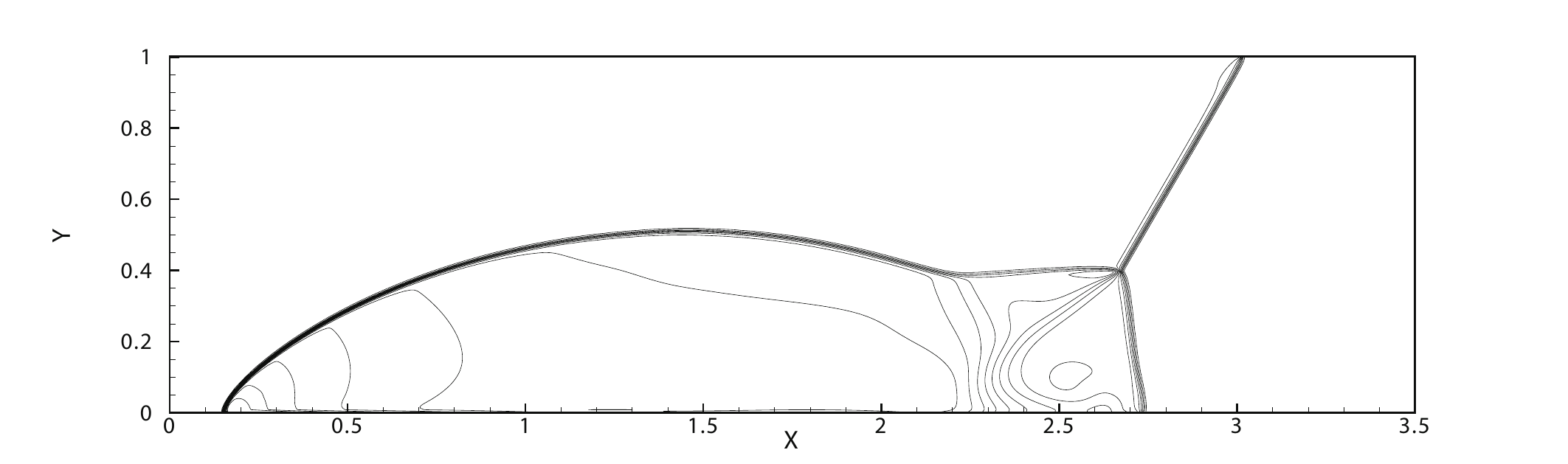}
		\end{minipage}
	}
	\caption{Density contours for double Mach reflection problem at $t = 0.2$.}
	\label{fig5.3}
\end{figure}

The computations are performed by first-order numerical schemes and the third-order TVD Runge-Kutta time discretization \cite{Shu1988} with CFL=0.5 up to t=0.2. The density contours computed by different schemes are shown in Fig. \ref{fig5.3}, where 20 contour levels varying from 2.0 to 20.0 are used. As shown, the HLLEM and HLLC schemes produce visible kinked Mach stems, demonstrating its vulnerability to shock instability. In contrast, the proposed ASHLLEM and ASHLLC solvers are both able to resolve shocks without any irregularities and the kinked Mach stems are barely noticeable.

\subsection{Hypersonic inviscid flow past a cylinder}
\label{S:5.4}
We continue to assess the shock robustness of all-speed HLL-type schemes by steady solutions of the hypersonic inviscid flow past a cylinder. As we all know, shock-capturing methods especially those with minimal dissipation on contact and shear waves are usually prone to shock anomalies such as the carbuncle phenomenon. A cylinder with a radius of the reference length is located in a uniform gas where the upstream Mach number is set as 20. The axis of the cylinder is at the origin $(x, y)= (0, 0)$. Since shock instabilities are sensitive to mesh systems, two mesh systems are used. These mesh systems are first proposed by Ohwada et al. \cite{Ohwada2013} to systemically assess shock robustness of kinetic schemes. One mesh system, Mesh-A, is defined by
\begin{equation}
\begin{aligned}
  \bar{x} = &  \left( {1 - \xi } \right)\left( {{a_1}\cosh \eta  - {a_2}} \right) - \xi \cos \eta, \\
  \bar{y} = &  {a_3}\left( {1 - \xi } \right)\sinh \eta  + \xi \sin \eta, \\
    {a_1} = & 2.45, \quad {a_2} = 4.736, \quad {a_3} = 3.185, \\
  \frac{1}{2} \leq & \xi\leq1, \quad -\frac{2\pi}{5}\leq\eta\leq\frac{2\pi}{5},
\end{aligned}
\label{sec5.4.1}
\end{equation}
where the spatial coordinates are normalized by the radius of the cylinder and the intervals for $\xi$ and $\eta$ are uniformly divided into $120$ and $320$ sections. This type of mesh is designed to make that grid lines around the shock wave align with it very well (not perfectly). The other mesh system, Mesh-B, is the cylindrical one defined by
\begin{equation}
\begin{aligned}
  \bar{x} = &  -(3.8-2.8\xi)\cos\eta, \\
  \bar{y} = &  (3.8-2.8\xi)\sin\eta, \\
\end{aligned}
\label{sec5.4.2}
\end{equation}
The intervals for $\xi$ and $\eta$ are the same as those in Mesh-A and they are uniformly divided as before. For this mesh, the uniform sections number $n_{\xi}\left(n_{\eta}\right)$ is chosen to create unite aspect ratio of cells around the shock wave. The computational domain has been initialized with values $\rho = 1.4$, $p=1$, $u=20$  and $v=0$. At the wall, the slip condition is used and the other two are taken as outflow. Simulations are conducted in first-order accurate schemes and the two-stage Runge-Kutta explicit time-stepping scheme with CFL = 0.5.
\begin{figure}[htbp]
	\centering
	\subfigure[$\rm{HLLEM}$]{
		\begin{minipage}[b]{0.20\textwidth}
			\centering
			\includegraphics[width=0.64\textwidth]{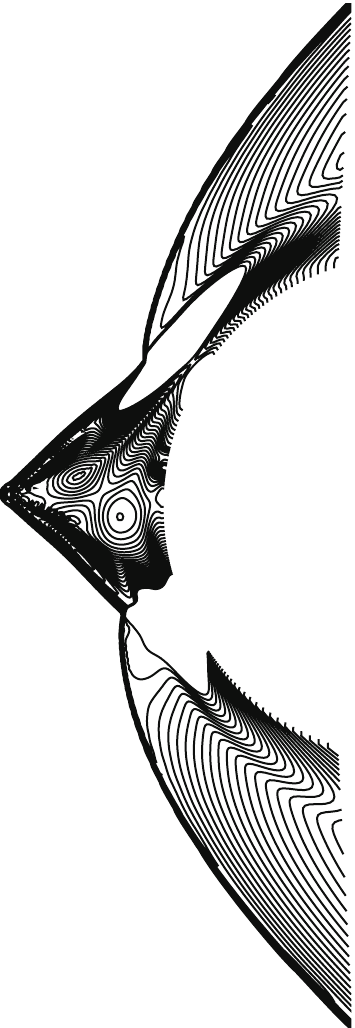}
		\end{minipage}
	}
	\subfigure[$\rm{HLLC}$]{
		\begin{minipage}[b]{0.20\textwidth}
			\centering
			\includegraphics[width=0.5\textwidth]{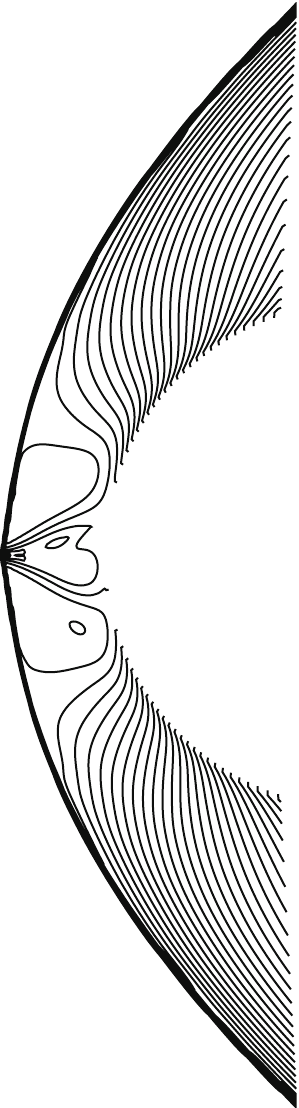}
		\end{minipage}
	}
	\subfigure[$\rm{ASHLLEM}$]{
		\begin{minipage}[b]{0.20\textwidth}
			\centering
			\includegraphics[width=0.5\textwidth]{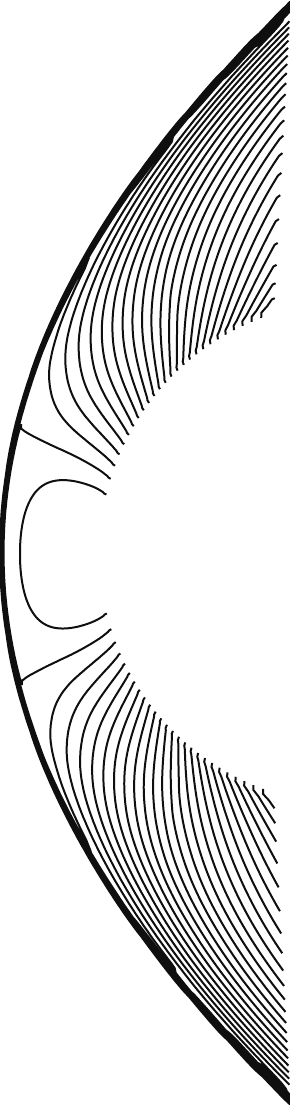}
		\end{minipage}
	}
	\subfigure[$\rm{ASHLLC}$]{
		\begin{minipage}[b]{0.20\textwidth}
			\centering
			\includegraphics[width=0.5\textwidth]{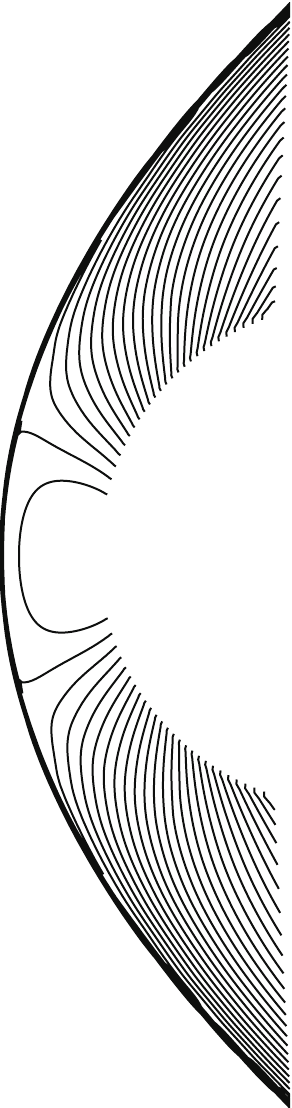}
		\end{minipage}
	}
	\caption{Comparison of density fields for different first-order accurate schemes on Mesh-A.}
	\label{fig5.4.1}
\end{figure}

\begin{figure}[htbp]
	\centering
	\subfigure[$\rm{HLLEM}$]{
		\begin{minipage}[b]{0.20\textwidth}
			\centering
			\includegraphics[width=1.10\textwidth]{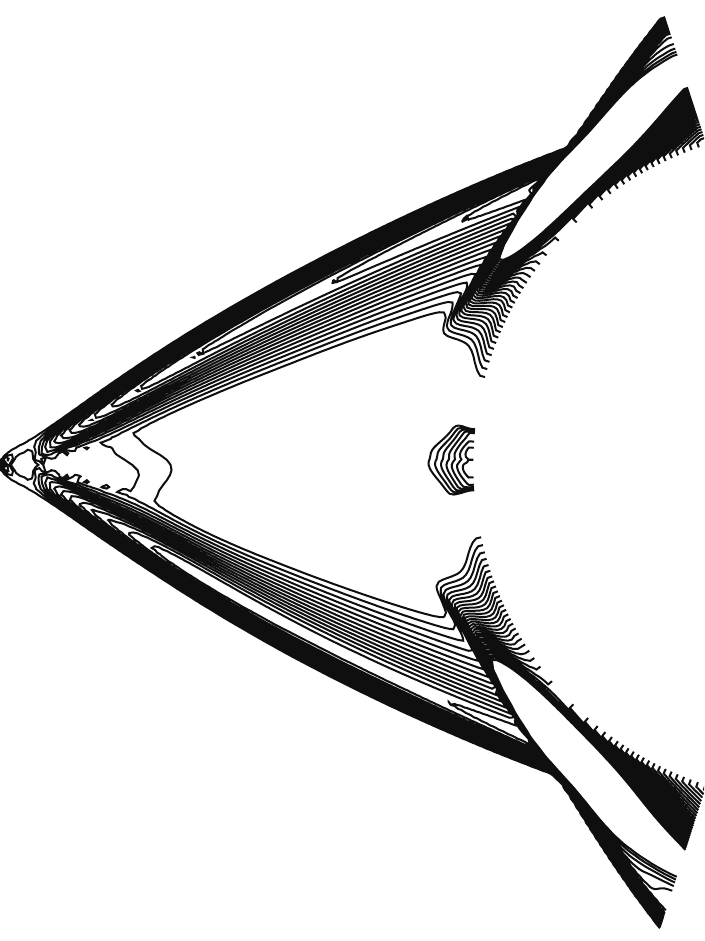}
		\end{minipage}
	}
	\subfigure[$\rm{HLLC}$]{
		\begin{minipage}[b]{0.20\textwidth}
			\centering
			\includegraphics[width=0.5\textwidth]{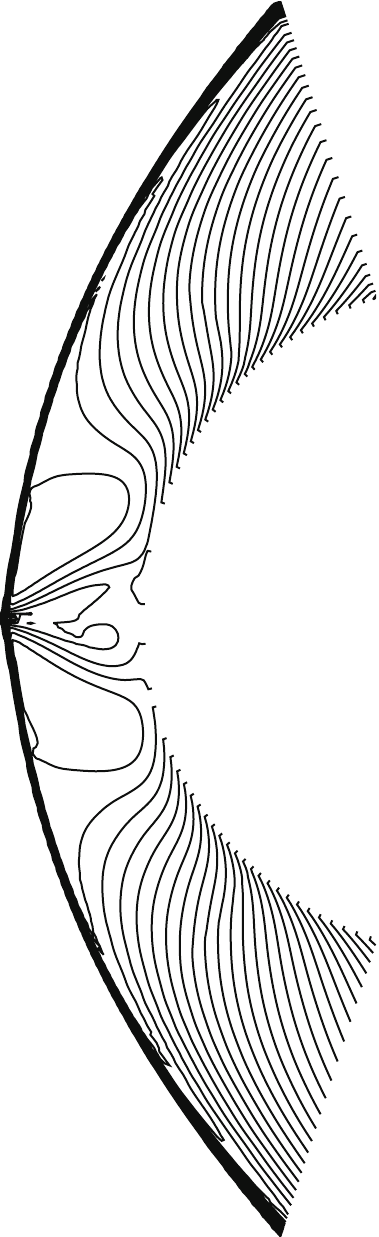}
		\end{minipage}
	}
	\subfigure[$\rm{ASHLLEM}$]{
		\begin{minipage}[b]{0.20\textwidth}
			\centering
			\includegraphics[width=0.5\textwidth]{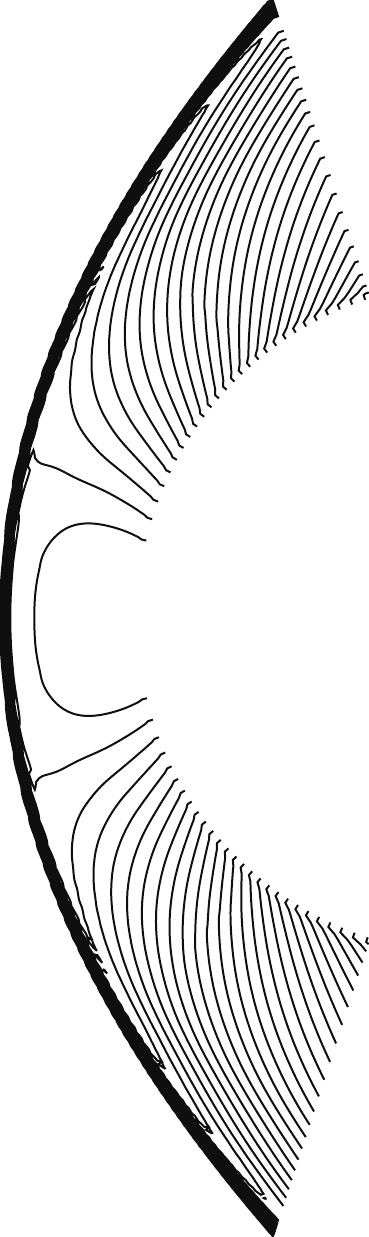}
		\end{minipage}
	}
	\subfigure[$\rm{ASHLLC}$]{
		\begin{minipage}[b]{0.20\textwidth}
			\centering
			\includegraphics[width=0.5\textwidth]{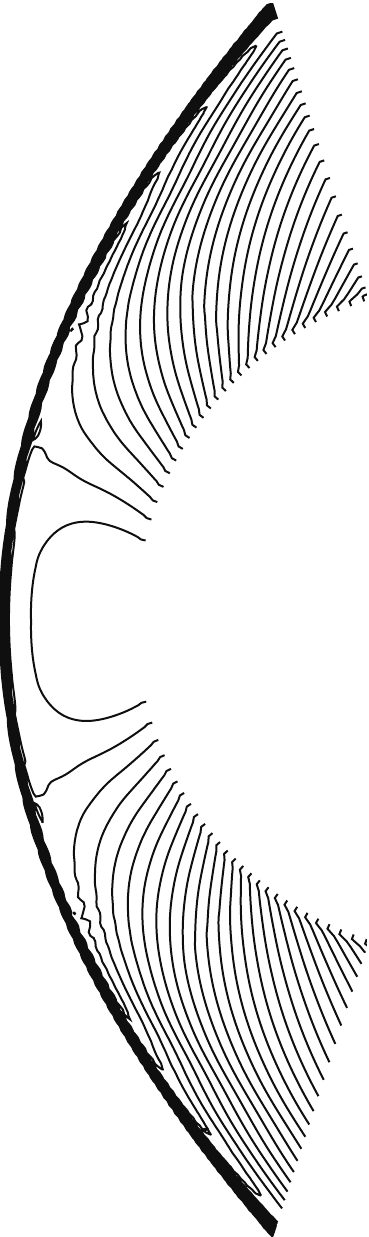}
		\end{minipage}
	}
	\caption{Comparison of density fields for different first-order accurate schemes on Mesh-B.}
	\label{fig5.4.2}
\end{figure}

\begin{figure}[htbp]
	\centering
	\subfigure[$\rm{ASHLLEM}$]{
		\begin{minipage}[b]{0.20\textwidth}
			\centering
			\includegraphics[width=0.5\textwidth]{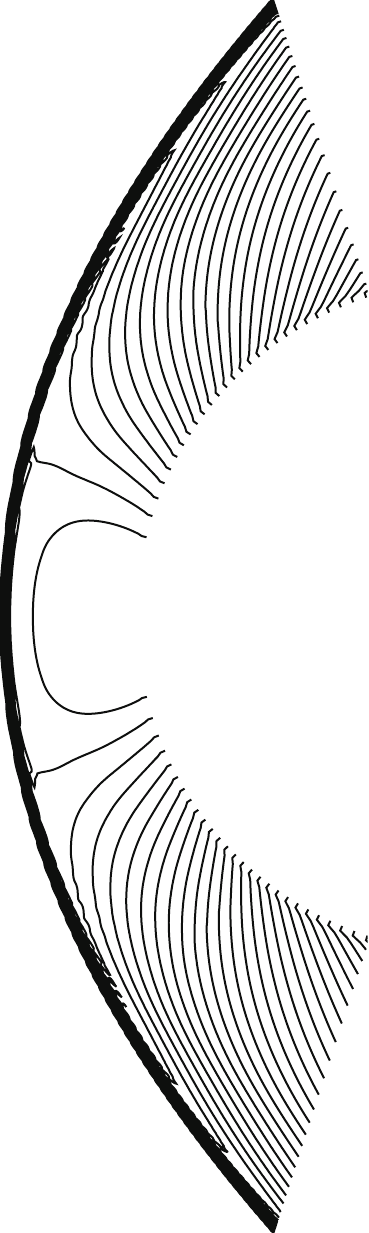}
		\end{minipage}
	}
	\subfigure[$\rm{ASHLLC}$]{
		\begin{minipage}[b]{0.20\textwidth}
			\centering
			\includegraphics[width=0.5\textwidth]{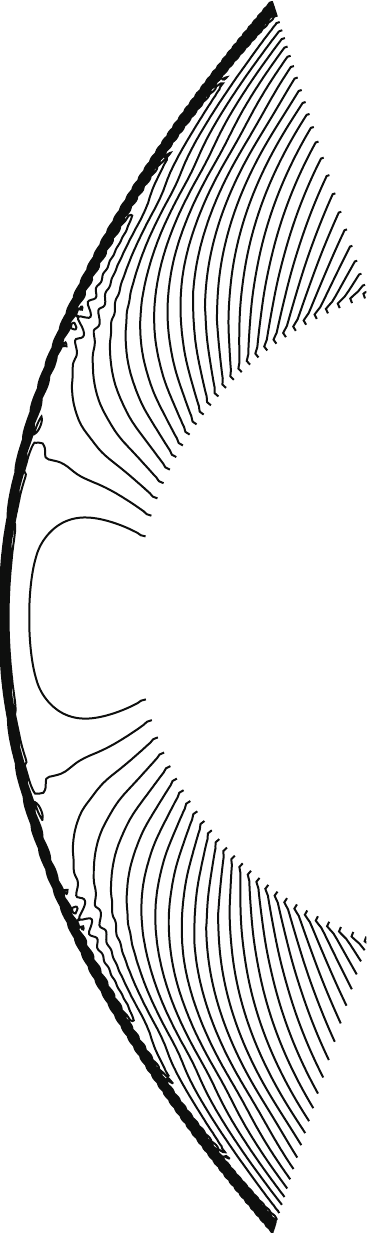}
		\end{minipage}
	}
	\caption{Comparison of density fields for first-order accurate all-speed schemes on Mesh-B (aspect ratio near the shock is 2.0).}
	\label{fig5.4.3}
\end{figure}

In Fig. \ref{fig5.4.1} and Fig. \ref{fig5.4.2}, density contours computed by different schemes on both mesh systems are illustrated, where 30 contour levels varying from 2.0 to 8.5 are used. As shown, original HLLEM and HLLC schemes exhibit the carbuncle phenomenon and appreciable post-shock wrinkles are visible. However, the proposed ASHLLEM and ASHLLC schemes produce clean shock profiles and their post shock regions are free from any shock anomalies. These computed results demonstrate that the proposed all-speed schemes are not only endowed with high resistance against strong shock waves, but also show a fairly high level of robustness in the case where the alignment of the computational mesh with the shock wave is poor. Furthermore, previous researches \cite{henderson2007grid,qu2019grid} have shown that the aspect ratio of cells near the shock is a major factor that influences the performances of shock-capturing methods against the shock instability. Cells with large aspect ratio are more prone to trigger shock instabilities near the shock wave. Thus, to further assess the robustness of all-speed HLL-type schemes for strong shock-capturing, the section number used to divide $\eta$ on Mesh-B is increased gradually. Computational results on these meshes show that the maximum aspect ratio of cells near the shock wave is about 2.0 for the ASHLLEM and ASHLLC schemes to resolve the shock wave stably. In Fig. \ref{fig5.4.3}, computational results for both methods are demonstrated, where nearly no visible shock anomalies and post-shock wrinkles appear.
%M20 the position of the shock is -1.39053296
\begin{figure}[htbp]
	\centering
	\subfigure{
		\begin{minipage}{4.0cm}
			\centering
			\includegraphics[scale=0.65]{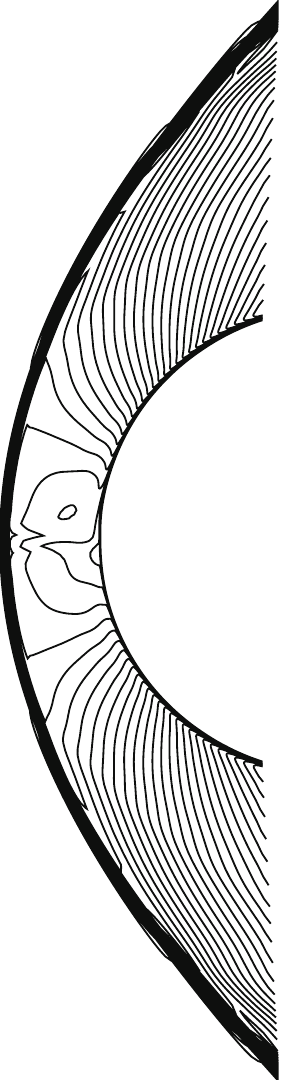}
		\end{minipage}
	}
	\subfigure{
		\begin{minipage}{4.0cm}
			\centering
			\includegraphics[scale=0.65]{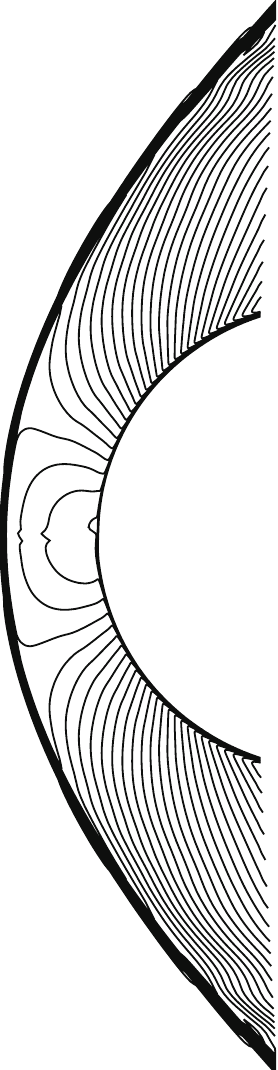}
		\end{minipage}
	}
	\subfigure{
		\begin{minipage}{4.0cm}
			\centering
			\includegraphics[scale=0.65]{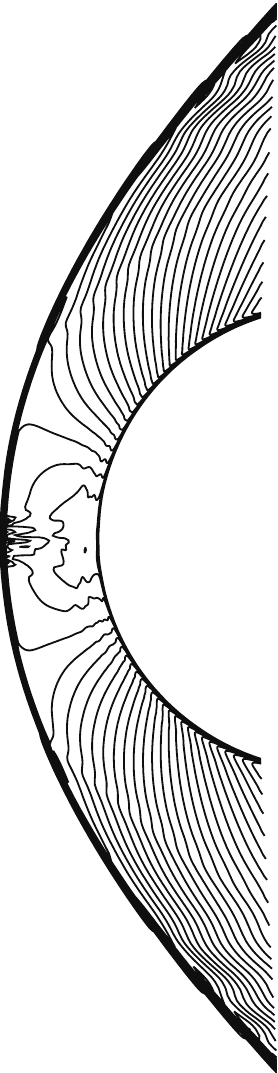}
		\end{minipage}
	}
	\subfigure{
		\begin{minipage}{4.0cm}
			\centering
			\includegraphics[scale=0.6]{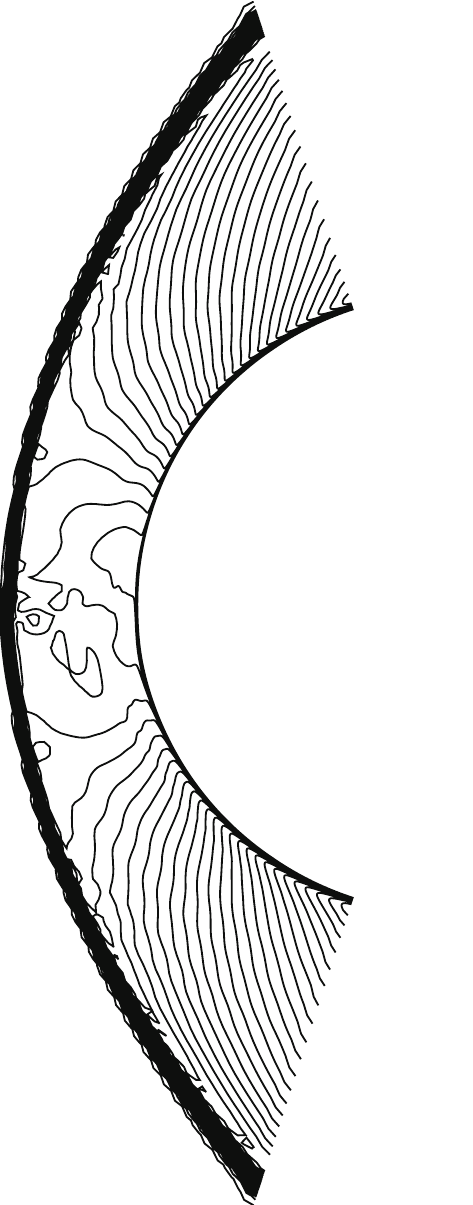}
		\end{minipage}
	}
	\subfigure{
		\begin{minipage}{4.0cm}
			\centering
			\includegraphics[scale=0.6]{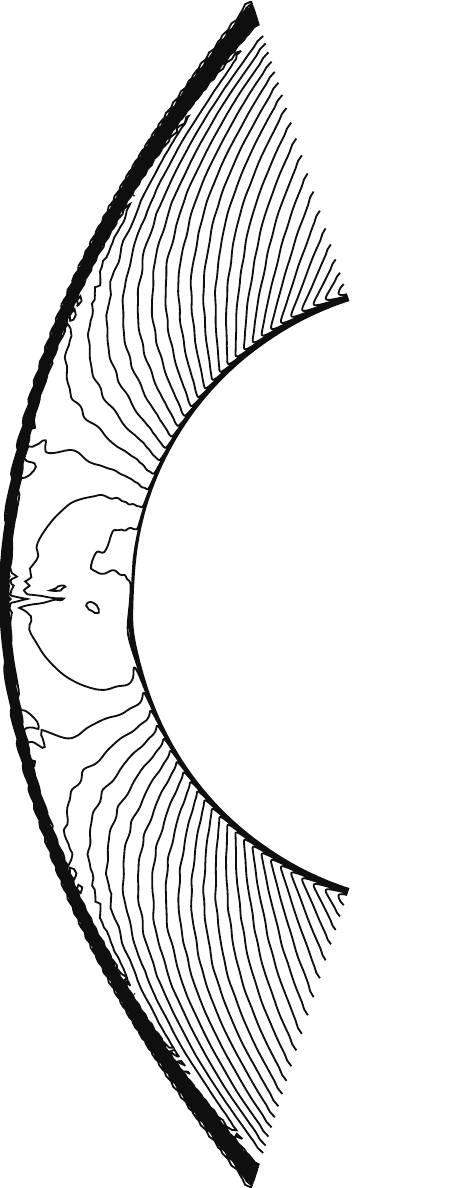}
		\end{minipage}
	}
	\subfigure{
		\begin{minipage}{4.0cm}
			\centering
			\includegraphics[scale=0.6]{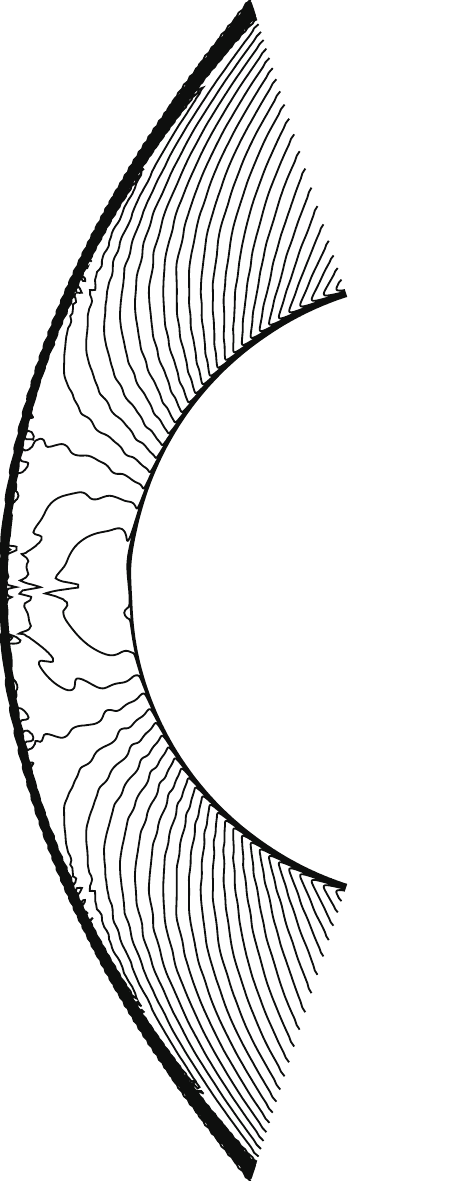}
		\end{minipage}
	}
	\caption{Comparison of density fields by HLLC scheme in Mach 8.1 viscous flow past a cylinder. $\left({n_{\xi}},{n_{\eta}}\right)=\left(120, 160\right)$; middle: $\left({n_{\xi}},{n_{\eta}}\right)=\left(180, 240\right)$; and right:$\left({n_ {\xi}},{n_{\eta}}\right)=\left(240, 320\right)$. The range of isolines is: $[1.05\le \rho /{\rho _{\inf }}\le6.9]$.}
	\label{fig5.5.1}
\end{figure}

\subsection{Hypersonic viscous flow past a cylinder}
\label{S:5.5}
The ultimate goal of the current study is to develop reliable and efficient shock-capturing schemes for hypersonic viscous flows especially the aeroheating problem. In this section, the problem of hypersonic viscous flow past a cylinder is used to examine the performance of all-speed HLL-type schemes for hypersonic heating computations. Here, the numerical setup follows that in references \cite{Ohwada2013,Kitamura2010}. The freestream conditions are given as $M_{\infty}=8.1$, $P_{\infty}=370.7 pa$, $T_{\infty}=63.73 K$ for the far field, and the Reynolds number based on the radius $\left(r=20mm\right)$ of the cylinder and the far field flow parameters is $1.3\times10^5$. The non-slip and isothermal conditions with the wall temperature $T_w=300K$ are imposed at the wall. The computational mesh adopted here is the same as that in the inviscid case, but the grids are refined near the cylinder surface to resolve the boundary layer well. Following Ohwada et al.\cite{Ohwada2013}, the non-uniform grid for $\xi$ is introduced in mesh-A and mesh-B defined in Eqs. ({\ref{sec5.4.1}}) and (\ref{sec5.4.2}) as
\begin{equation}
  \xi=\frac{81-41{\exp}(-s)}{80} \quad 0 \le s \le ln(41),
\end{equation}
and the interval for $s$ is divided into $n_{\xi}$ uniform sections.

\begin{figure}[htbp]
	\centering
	\subfigure{
		\begin{minipage}{4.0cm}
			\centering
			\includegraphics[scale=0.6]{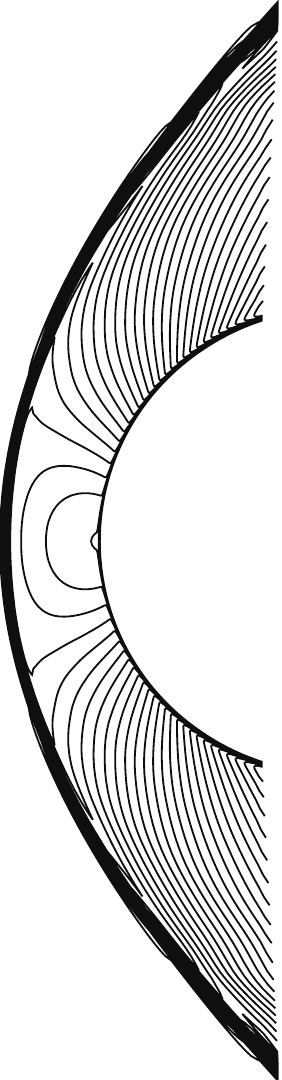}
		\end{minipage}
	}
	\subfigure{
		\begin{minipage}{4.0cm}
			\centering
			\includegraphics[scale=0.6]{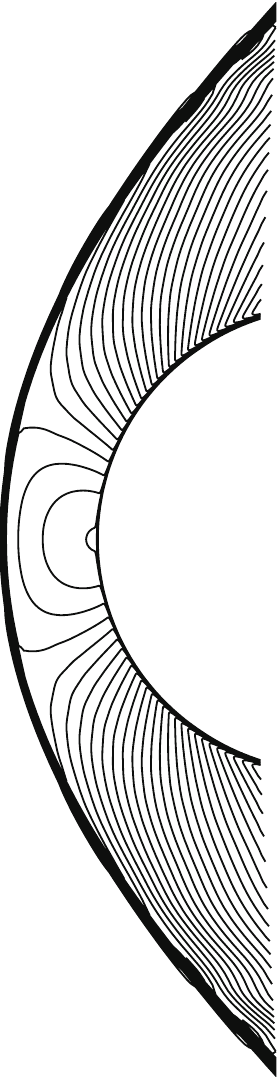}
		\end{minipage}
	}
	\subfigure{
		\begin{minipage}{4.0cm}
			\centering
			\includegraphics[scale=0.6]{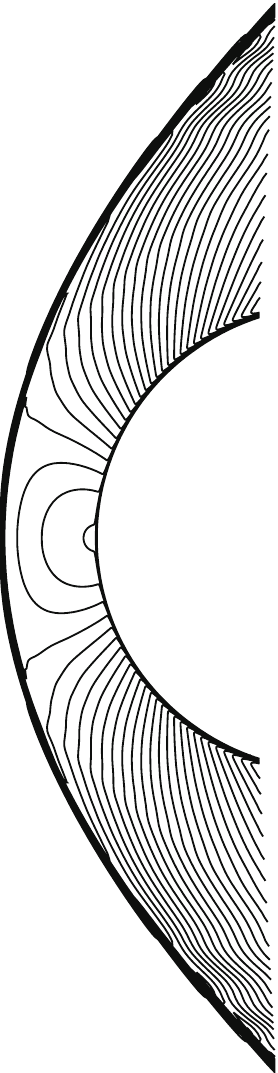}
		\end{minipage}
	}
	\subfigure{
		\begin{minipage}{4.0cm}
			\centering
			\includegraphics[scale=0.6]{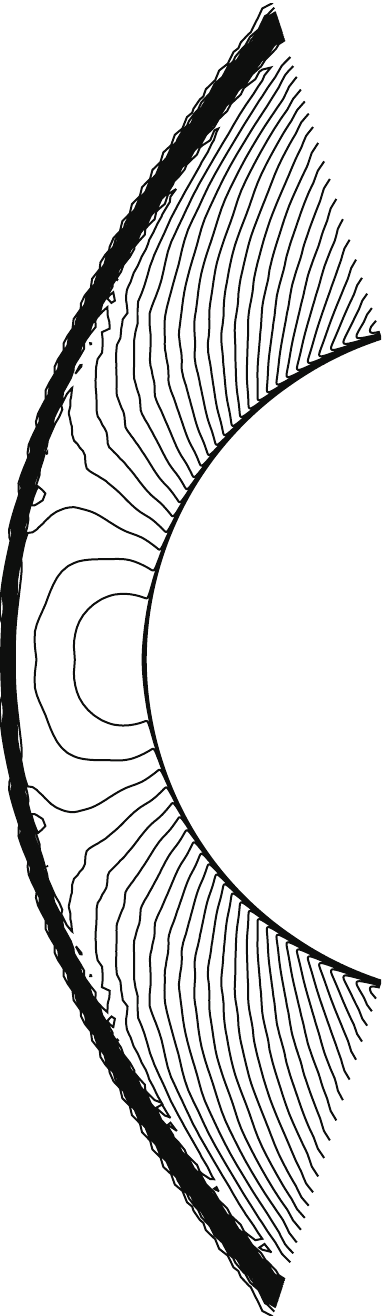}
		\end{minipage}
	}
	\subfigure{
		\begin{minipage}{4.0cm}
			\centering
			\includegraphics[scale=0.6]{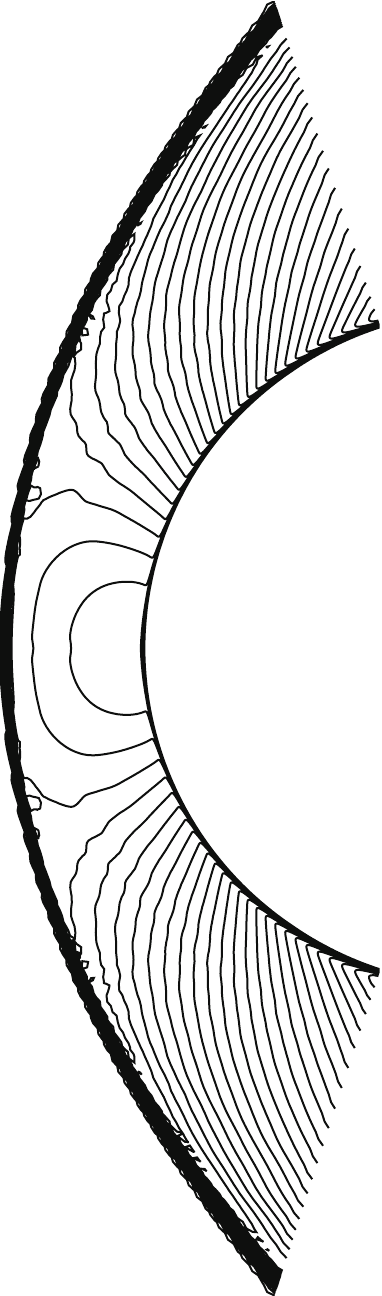}
		\end{minipage}
	}
	\subfigure{
		\begin{minipage}{4.0cm}
			\centering
			\includegraphics[scale=0.6]{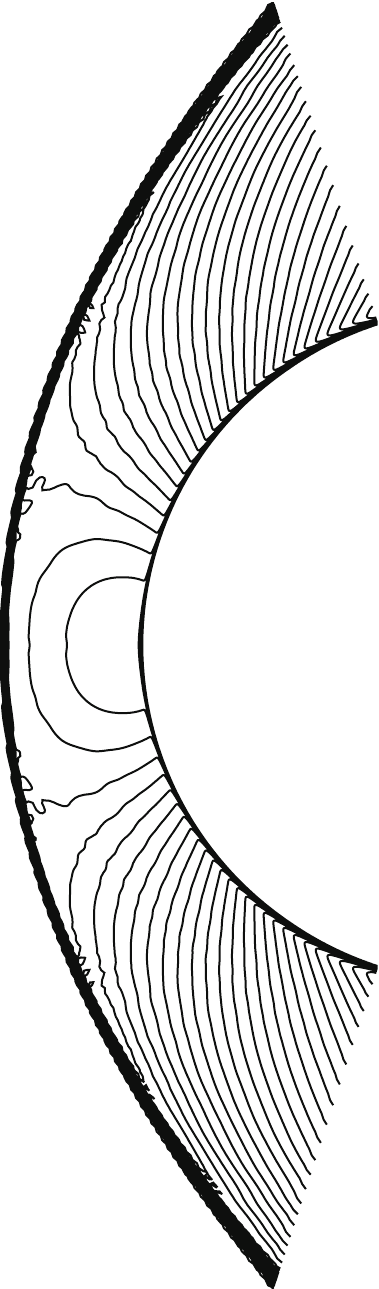}
		\end{minipage}
	}
	\caption{Comparison of density fields by ASHLLEM scheme in Mach 8.1 viscous flow past a cylinder. $\left({n_{\xi}},{n_{\eta}}\right)=\left(120, 160\right)$; middle: $\left({n_{\xi}},{n_{\eta}}\right)=\left(180, 240\right)$; and right:$\left({n_ {\xi}},{n_{\eta}}\right)=\left(240, 320\right)$. The range of isolines is: $[1.05\le \rho /{\rho _{\inf }}\le6.9]$.}
	\label{fig5.5.2}
\end{figure}

\begin{figure}[htbp]
	\centering
	\subfigure{
		\begin{minipage}{4.0cm}
			\centering
			\includegraphics[scale=0.65]{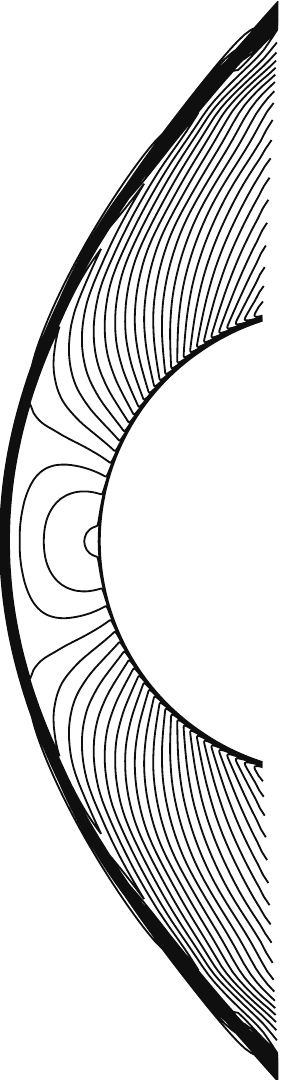}
		\end{minipage}
	}
	\subfigure{
		\begin{minipage}{4.0cm}
			\centering
			\includegraphics[scale=0.65]{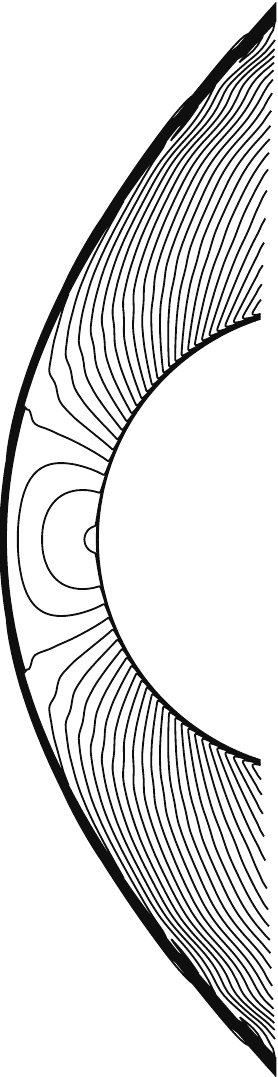}
		\end{minipage}
	}
	\subfigure{
		\begin{minipage}{4.0cm}
			\centering
			\includegraphics[scale=0.65]{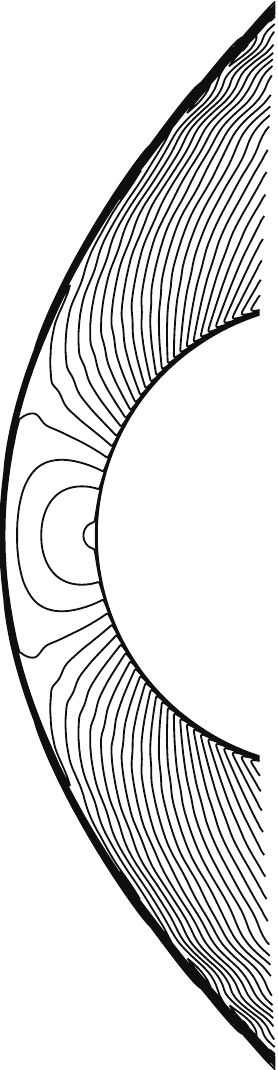}
		\end{minipage}
	}
	\subfigure{
		\begin{minipage}{4.0cm}
			\centering
			\includegraphics[scale=0.6]{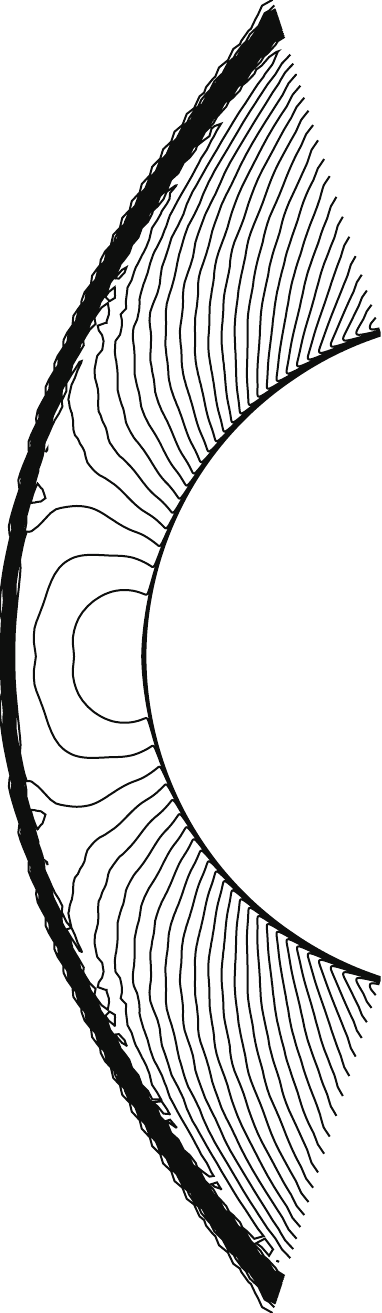}
		\end{minipage}
	}
	\subfigure{
		\begin{minipage}{4.0cm}
			\centering
			\includegraphics[scale=0.6]{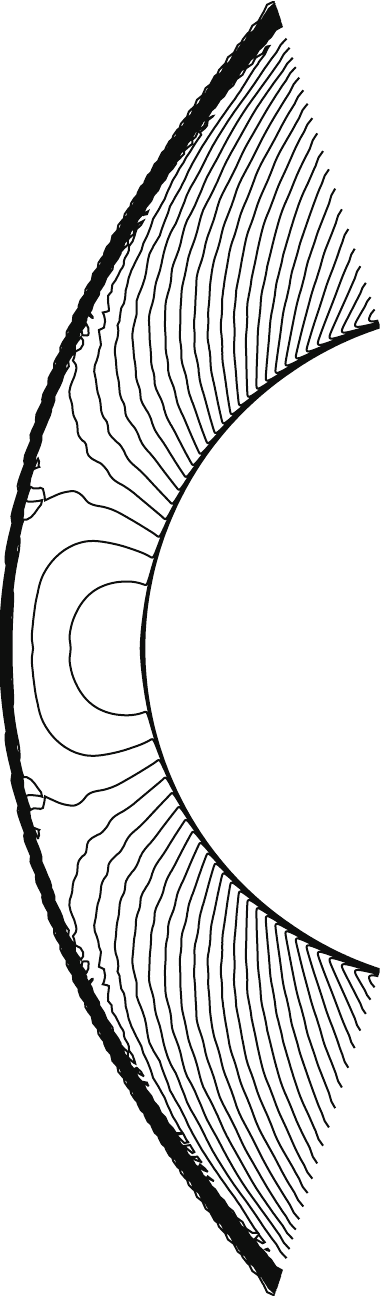}
		\end{minipage}
	}
	\subfigure{
		\begin{minipage}{4.0cm}
			\centering
			\includegraphics[scale=0.6]{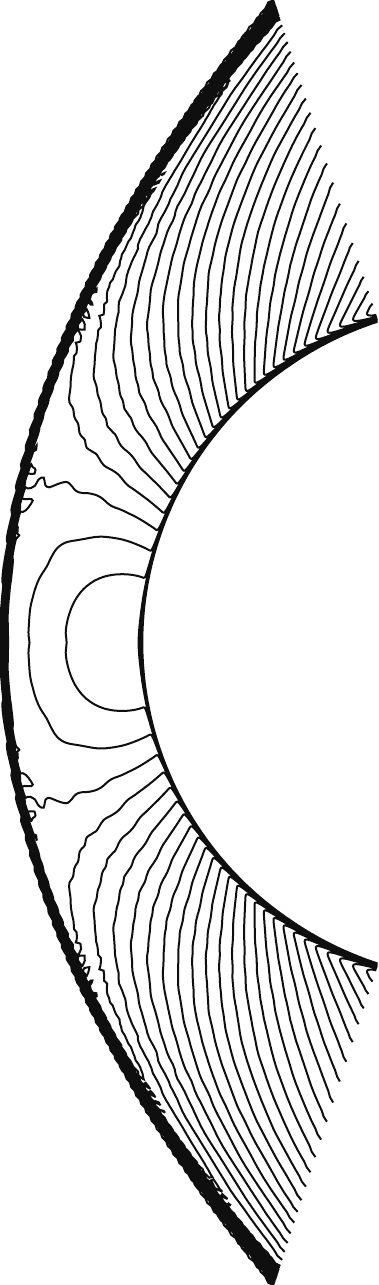}
		\end{minipage}
	}
	\caption{Comparison of density fields by ASHLLC scheme in Mach 8.1 viscous flow past a cylinder. $\left({n_{\xi}},{n_{\eta}}\right)=\left(120, 160\right)$; middle: $\left({n_{\xi}},{n_{\eta}}\right)=\left(180, 240\right)$; and right:$\left({n_ {\xi}},{n_{\eta}}\right)=\left(240, 320\right)$. The range of isolines is: $[1.05\le \rho /{\rho _{\inf }}\le6.9]$.}
	\label{fig5.5.3}
\end{figure}

\begin{figure}[htbp]
  \centering
  \subfigure[HLLC]{
  \includegraphics[width=0.48\textwidth]{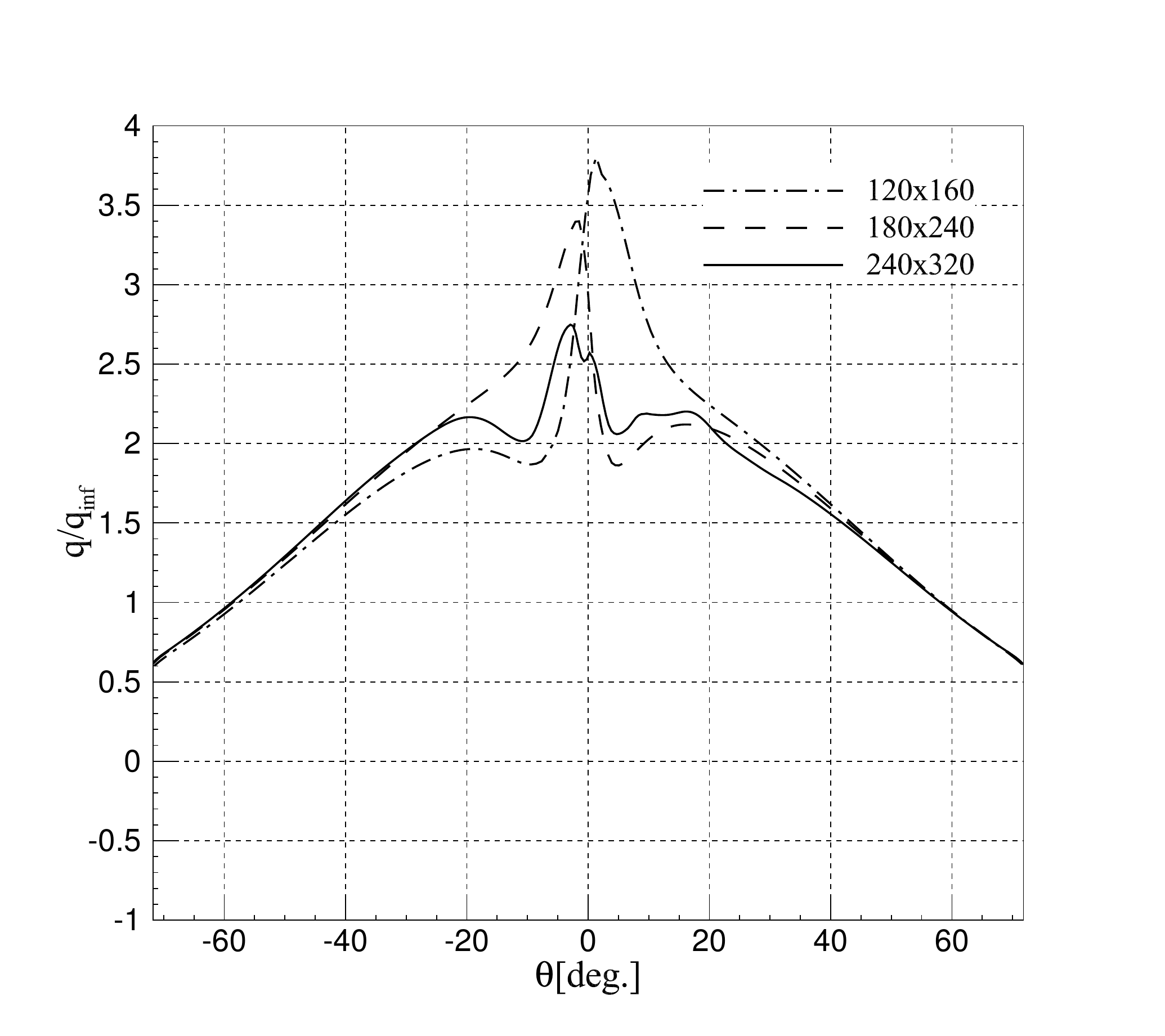}}
  \quad
  \subfigure[ASHLLEM]{
  \includegraphics[width=0.48\textwidth]{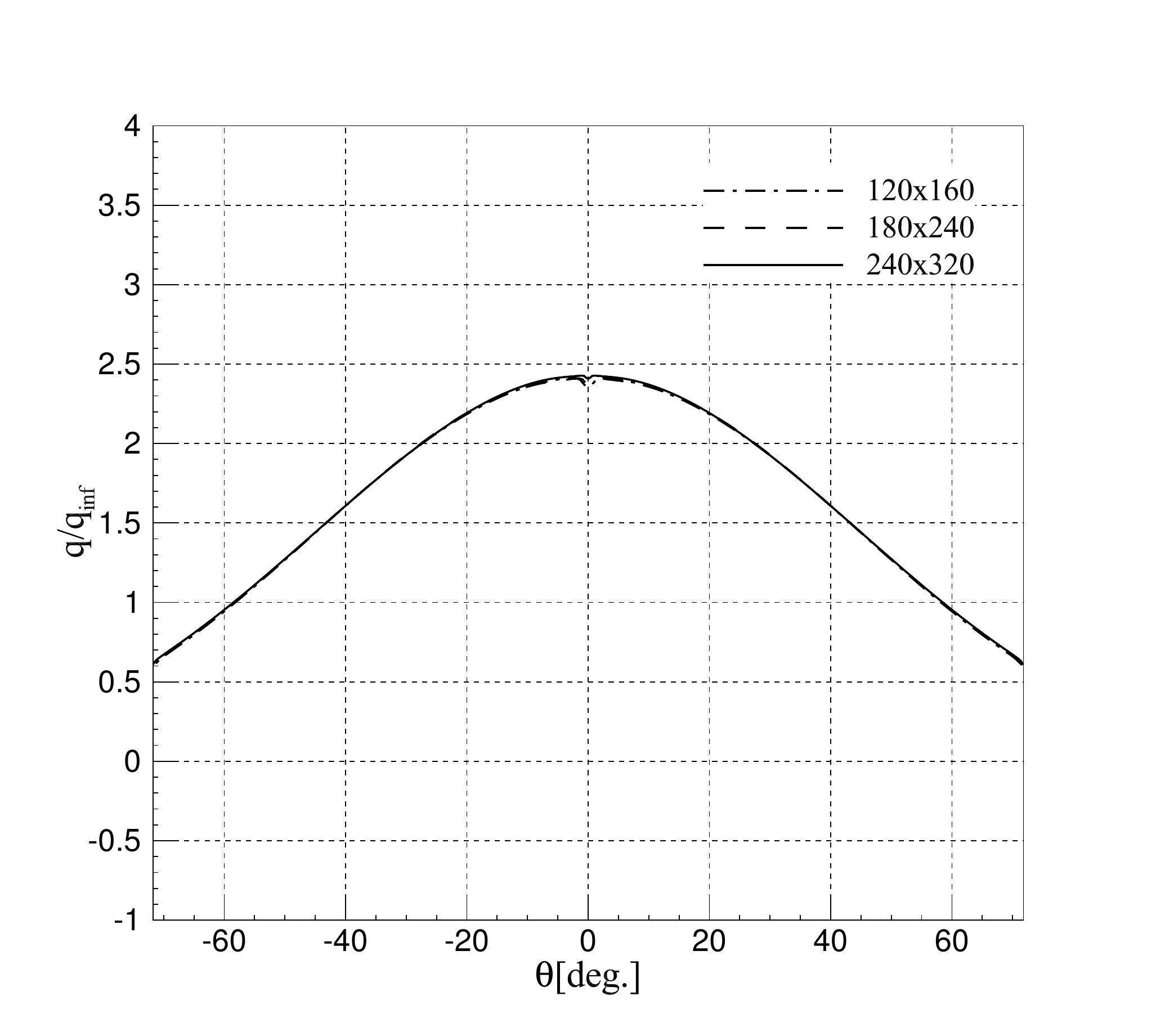}}
  \subfigure[ASHLLC]{
  \includegraphics[width=0.48\textwidth]{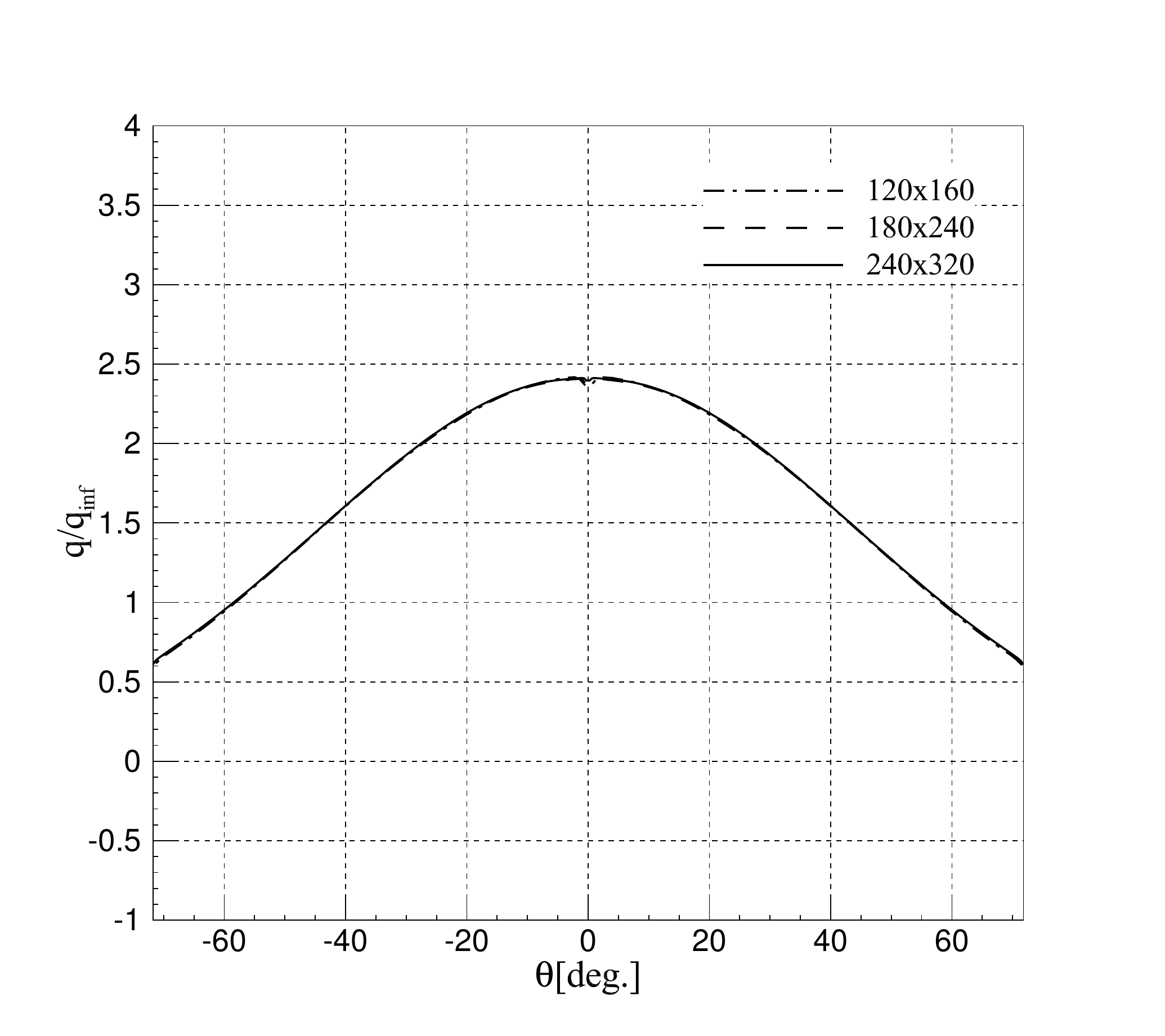}}
  \caption{Heat flux distribution along the cylinder surface, Mesh A.}
  \label{fig5.5.4}
\end{figure}

\begin{figure}[htbp]
  \centering
  \subfigure[HLLC]{
  \includegraphics[width=0.48\textwidth]{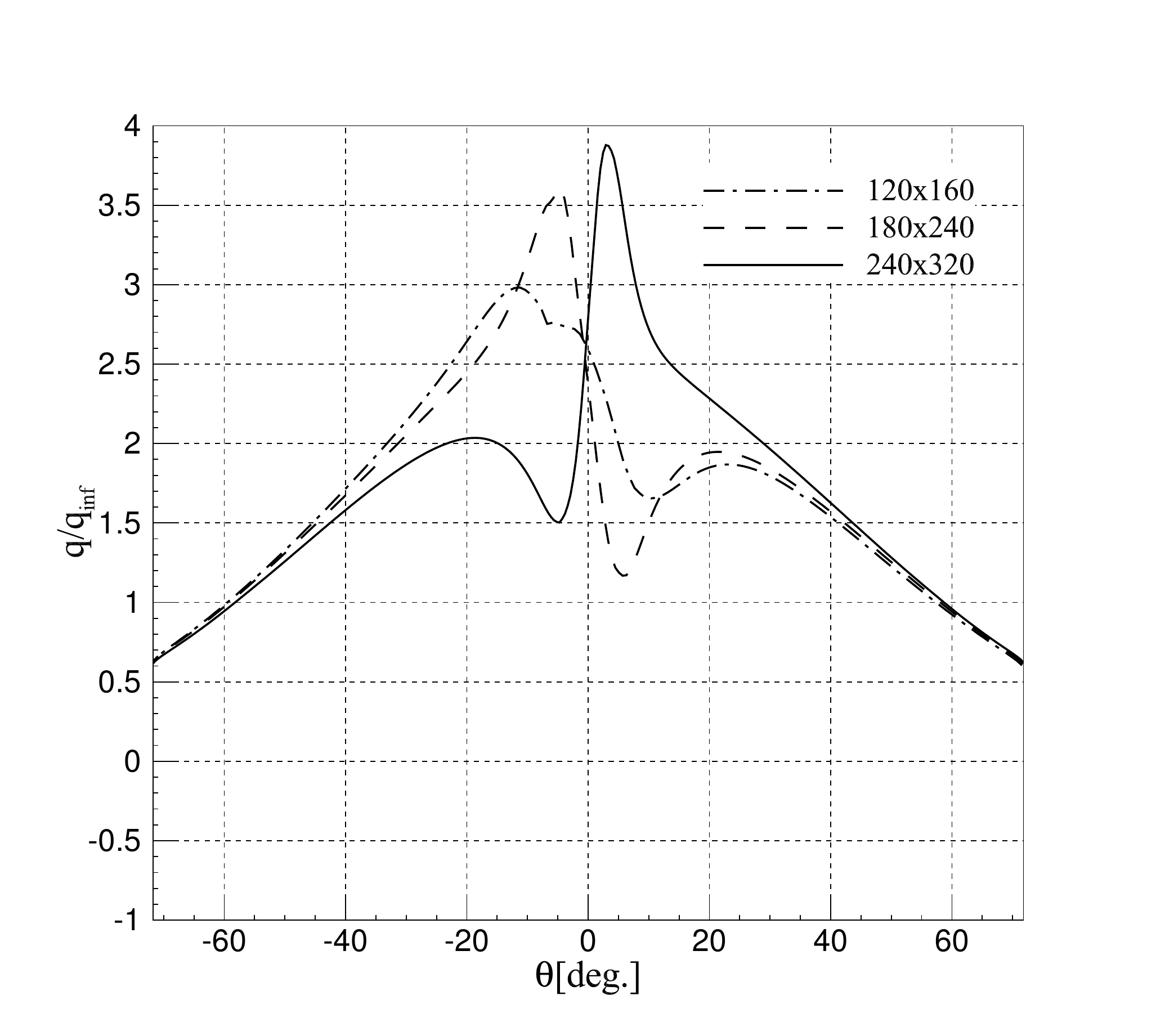}}
  \quad
  \subfigure[ASHLLEM]{
  \includegraphics[width=0.48\textwidth]{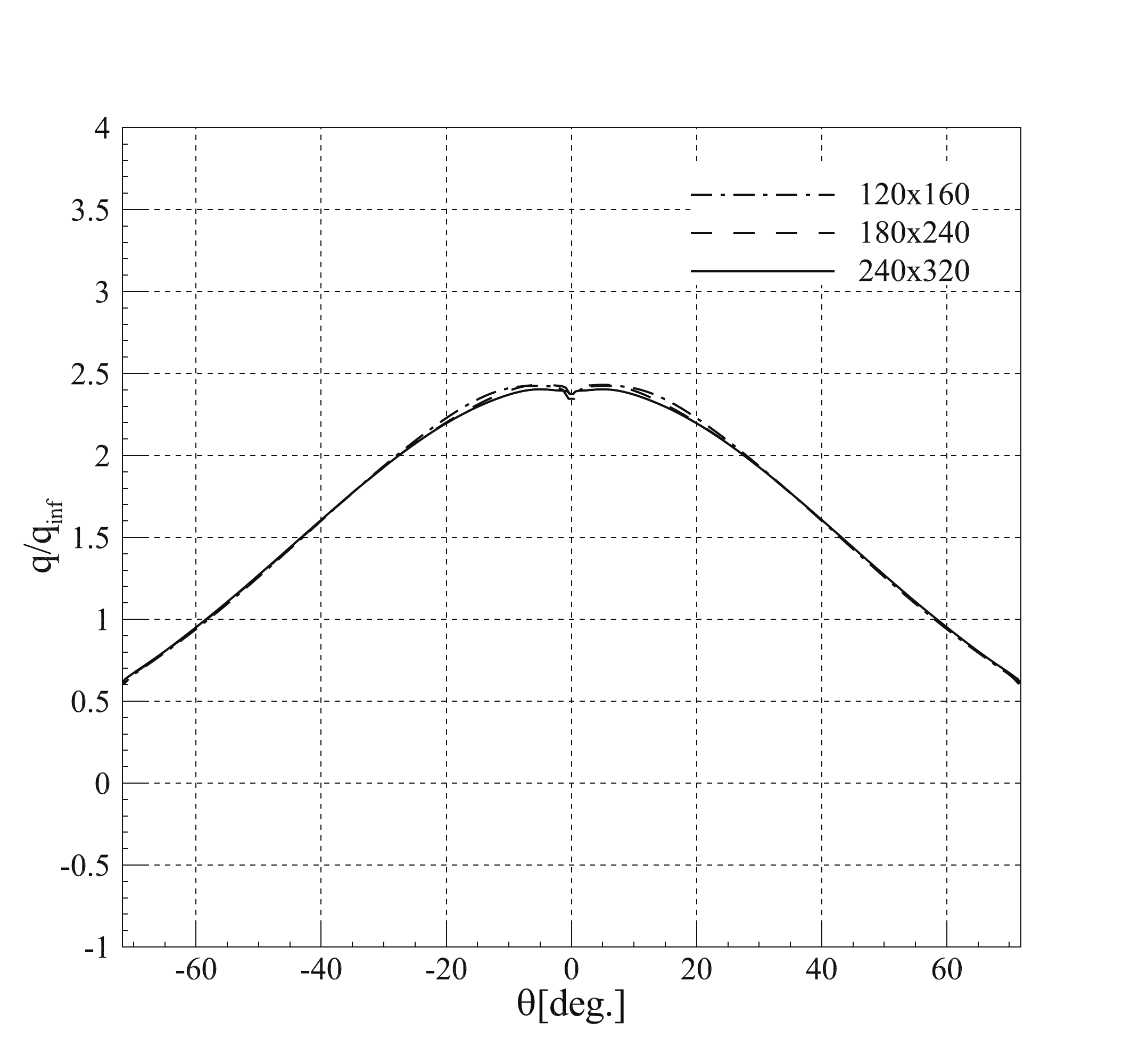}}
  \subfigure[ASHLLC]{
  \includegraphics[width=0.48\textwidth]{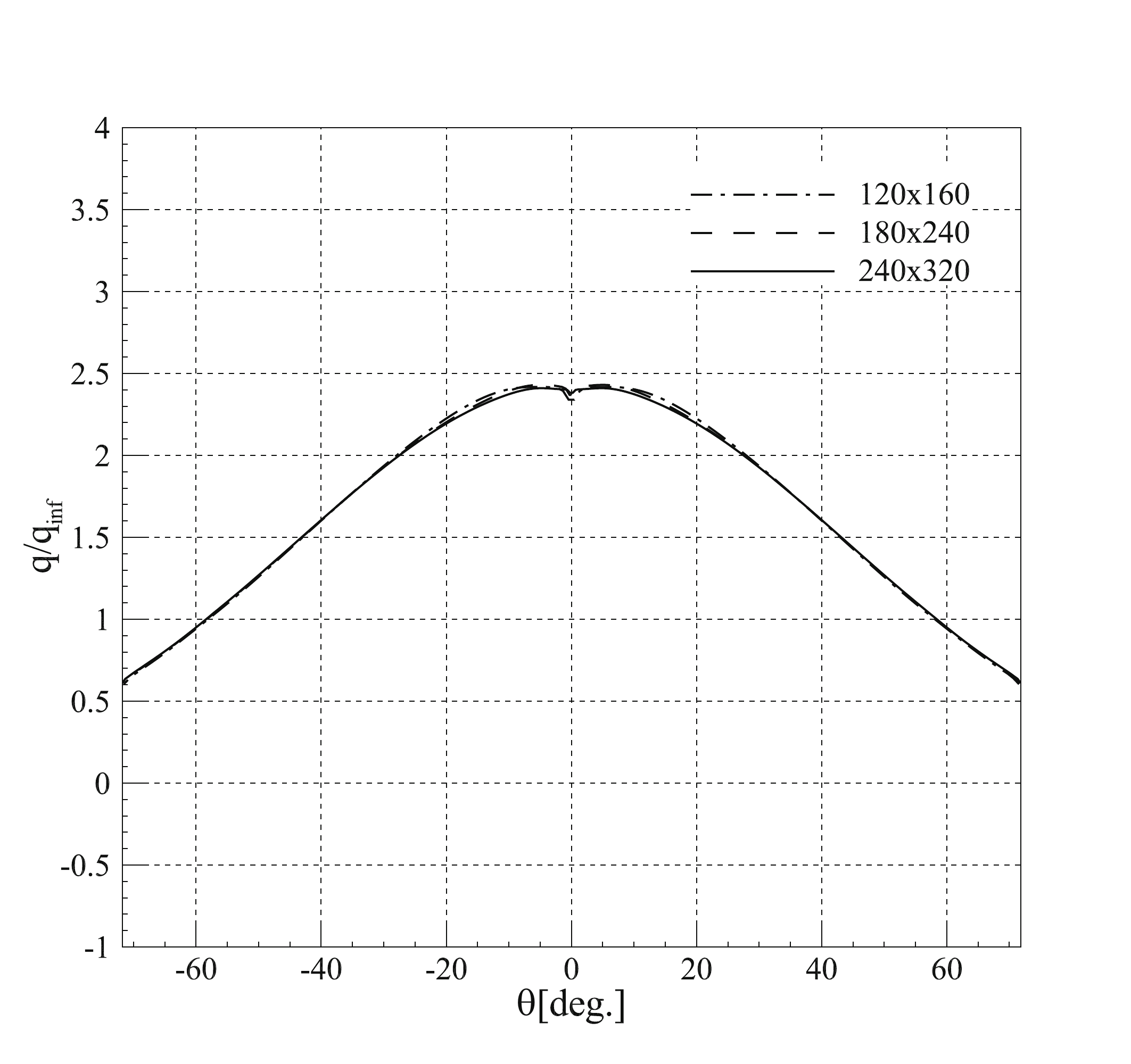}}
  \caption{Heat flux distribution along the cylinder surface, Mesh B.}
  \label{fig5.5.5}
\end{figure}

The second-order-accurate, van Albada-limited MUSCL reconstruction ($\kappa=1/3$) is adopted for cell-interface values, along with the second-order central difference for viscous term.
For temporal discretization, the LU-SGS approach is employed. All the computations are conducted for 100,000 steps with $\rm{CFL}=200$. The residuals defined as the L2-norm of density drop at least three orders of magnitude for all the cases. In Fig. \ref{fig5.5.1}$\sim$\ref{fig5.5.3}, the density fields of different flux functions for three meshes $\left(n_{\xi}, n_{\eta}\right)=(120, 160), (180, 240), (240, 320)$ are depicted, where the HLLEM scheme fails the computation and its results are not shown. As shown, the density fields computed by the HLLC scheme exhibit shock anomalies, i.e. the carbuncle phenomenon and the post-shock wrinkles. Whereas, the proposed ASHLLEM and ASHLLC schemes still produce clear and symmetrical density fields and shock anomalies are barely visible. In Fig. \ref{fig5.5.4} and Fig. \ref{fig5.5.5}, we provide the profiles of the nondimensional heat flux $q/q_{inf}$ along the cylinder surface for different schemes. As shown, the results computed by the HLLC scheme are mesh dependent and very inaccurate due to shock anomalies. Due to the failure of HLLEM scheme for the computation, its heat flux distributions are not shown. However, the results of the all-speed schemes are nearly mesh independent, although small pimple-like variation is observed around $\theta {\rm{ = }}0$ in some cases. Moreover, the value of $q/q_{inf}$ at the stagnation point computed by ASHLLEM and ASHLLC schemes is in good agreement with the theoretical value $q/q_{inf}=2.46$, which is predicted by Fay-Riddell \cite{Fay1958}. The above results demonstrate that the proposed ASHLLEM and ASHLLC schemes are able to produce accurate and reliable results for hypersonic heating computations.

\subsection{Hypersonic viscous flow past a space shuttle}
\label{S:5.6}
\begin{figure}[htbp]
  \centering
  \includegraphics[width=0.48\textwidth]{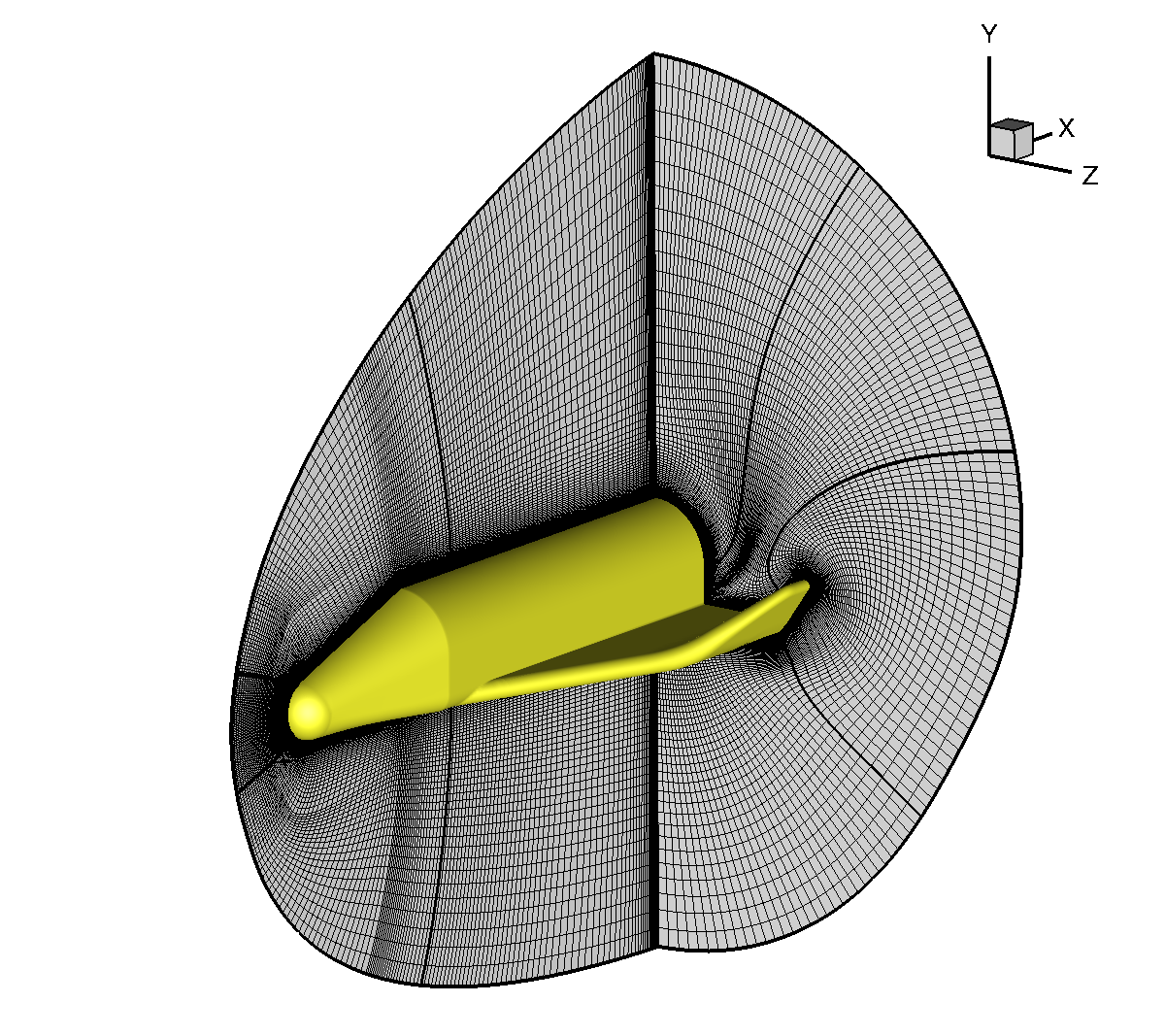}
  \caption{Computational grids over the space shuttle model.}
  \label{fig5.6.1}
\end{figure}

In this last test case, we assess the performance of the proposed all-speed HLL-type schemes for calculations of complex three-dimensional configurations. The problem considered here is the hypersonic viscous flow past a space shuttle. It has detailed experimental data \cite{Lisuxun} and can be used to access the performance of proposed schemes for the calculation of heat flux. The geometry of the space shuttle and its corresponding computational grid are presented in Fig. \ref{fig5.6.1}, where a total number of 3.56 million nonoverlapping hexahedral grids are used to discrete the computational domain. The minimal grid interval in the normal direction near the wall is $1.0\times10^{-5}$. The computational conditions are set as the same as the corresponding wind-tunnel test, i.e., $M_{\infty}=10.02$, $Re_{\infty}=2.2\times10^6$, $T_{\infty}=69.12$, $T_{w}=294.4$ and $\alpha=0$. Both all-speed HLL-type schemes are used with MUSCL reconstruction and van Albada limiter \cite{vanAlbada1997} to solve the Navier-Stokes equations. An implicit LU-SGS approach with CFL=5 for 100,000 time iterations is used to carried out the simulation.

\begin{figure}[htbp]
  \centering
  \subfigure[ASHLLEM]{
  \includegraphics[width=0.48\textwidth]{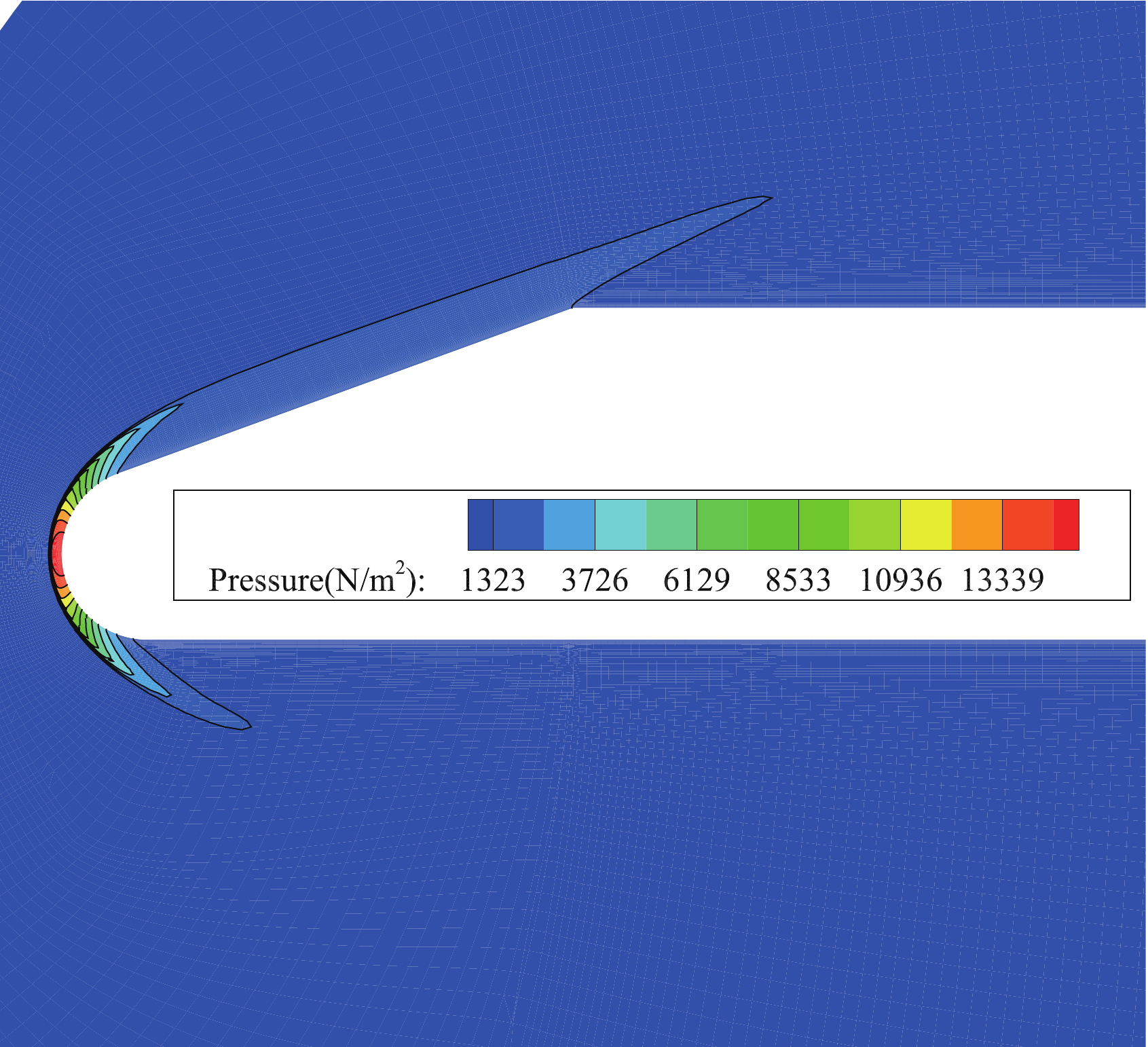}}
  \subfigure[ASHLLC]{
  \includegraphics[width=0.48\textwidth]{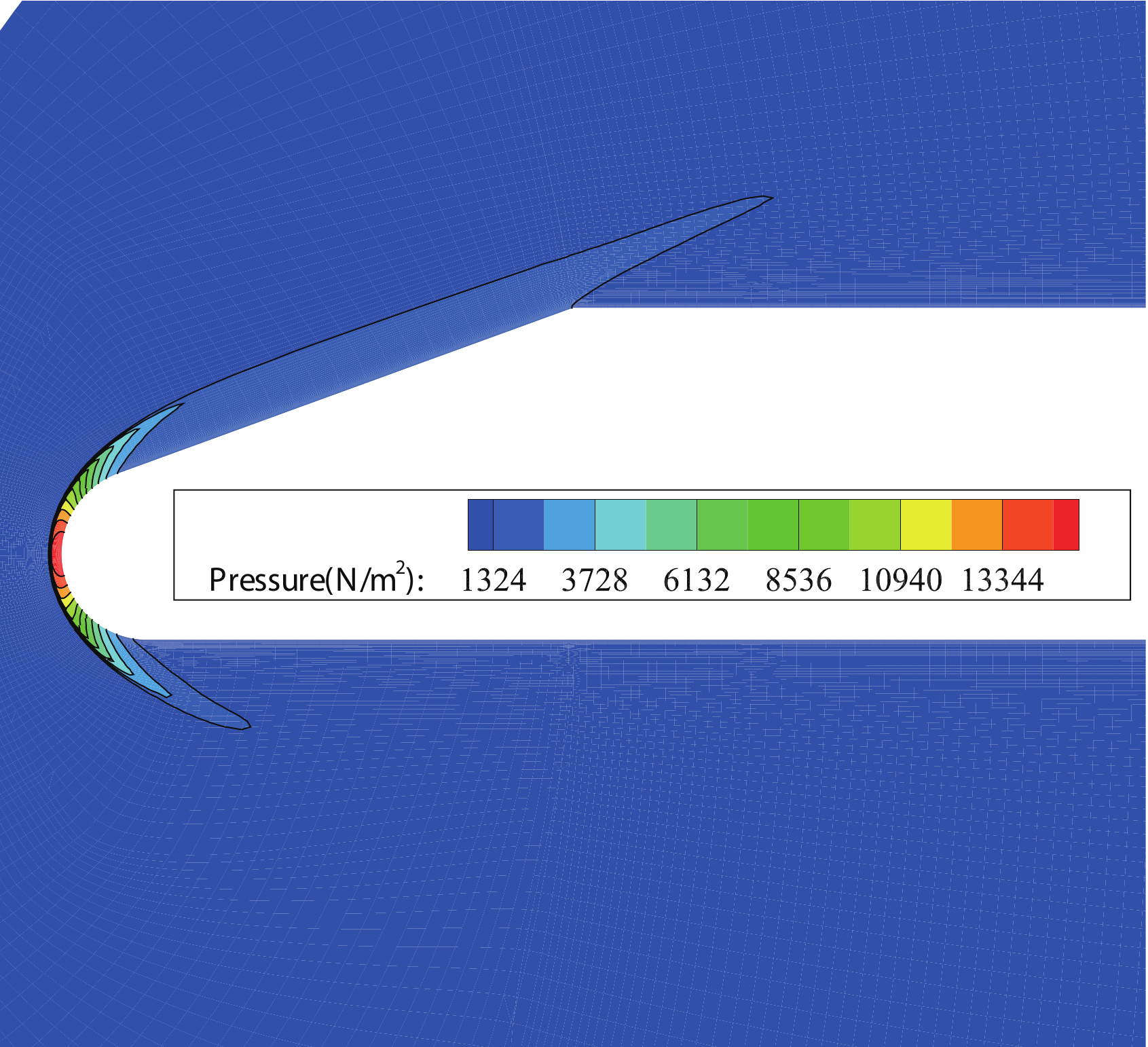}}
  \caption{Pressure contours on the symmetry plane.}
  \label{fig5.6.2}
\end{figure}

\begin{figure}[htbp]
  \centering
  \subfigure[windward]{
  \includegraphics[width=0.48\textwidth]{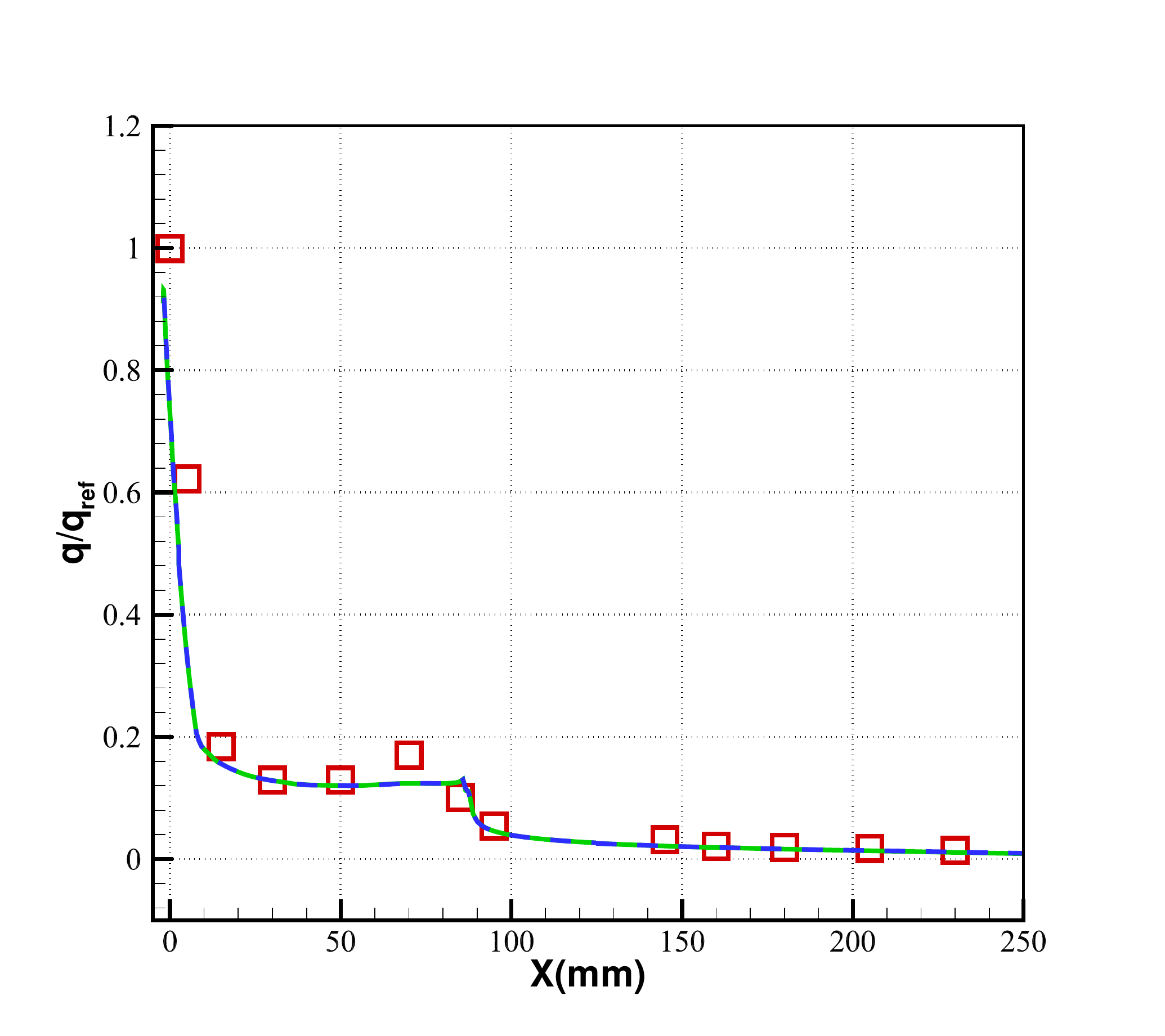}}
  \subfigure[leeward]{
  \includegraphics[width=0.48\textwidth]{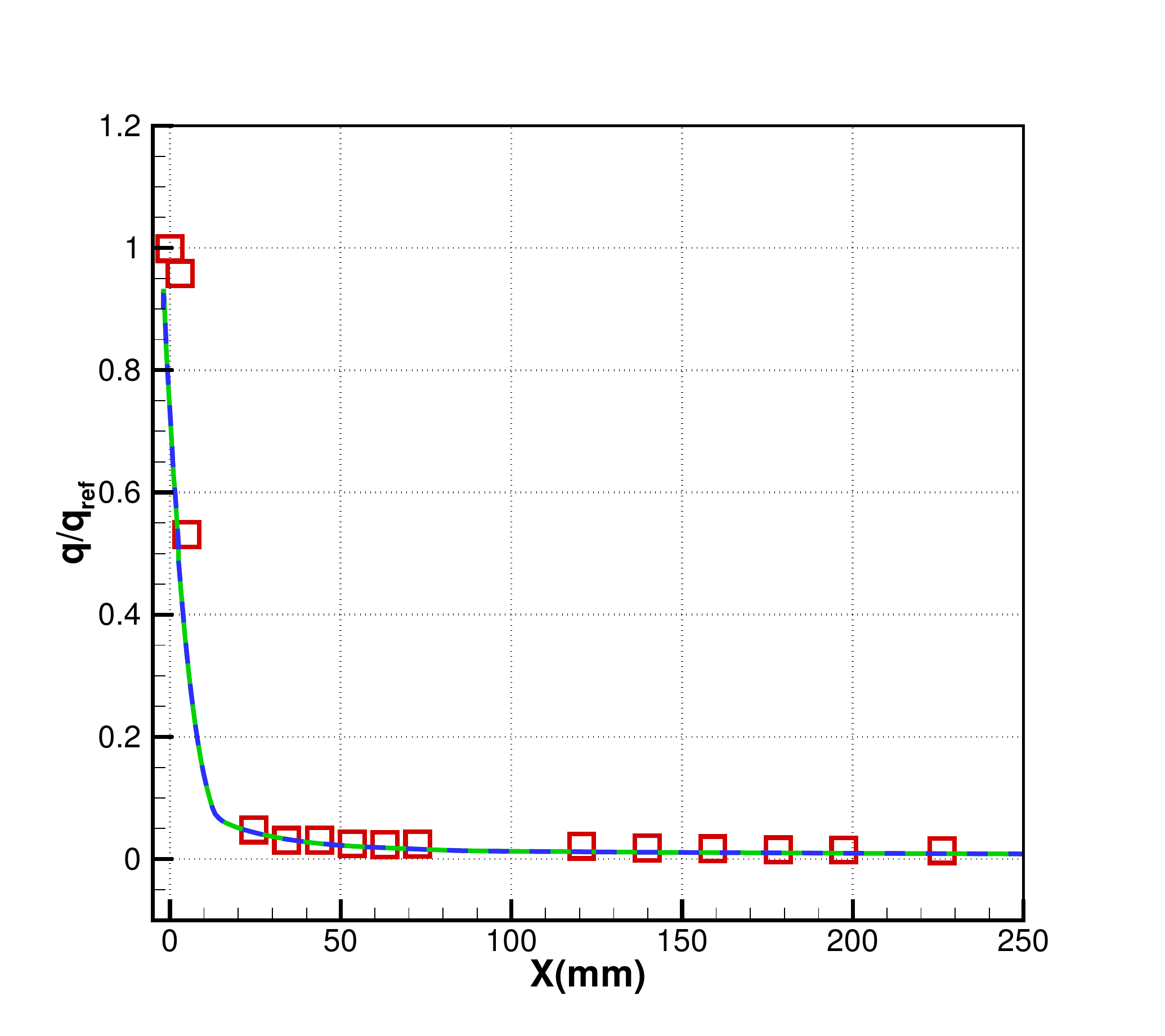}}
  \caption{Heat flux distribution along the symmetric lines. Experimental data (red square), ASHLLEM (green solid line) and ASHLLC (blue dashed).}
  \label{fig5.6.3}
\end{figure}
The computational results are demonstrated in Fig. \ref{fig5.6.2}, in which the pressure contours on the symmetry plane are indicated. As can be seen from this figure, the bow shocks ahead of the model nose are well-captured by both ASHLLEM and ASHLLC schemes without visible oscillations. Fig. \ref{fig5.6.3} demonstrates the heat distributions on the windward and leeward symmetric lines, in which the computational heat flux agree well with the experimental data \cite{Lisuxun}. All the computational results demonstrate the capability of the present all-speed schemes for applications of the three-dimensional hypersonic vehicle.

\section{Conclusions}
\label{S:6}
In the current study, we devote our efforts to developing a general approach to construct all-speed HLL-type schemes for reliable computations of hypersonic heating problems. The proposed two all-speed HLL-type schemes called ASHLLEM and ASHLLC are not only endowed with high resistance against shock anomalies, but also enjoy the property of low dissipation at low speeds. This is implemented by modifying the original HLLEM and HLLC flux functions with a shock stabilization technique and a low-Mach fix method in the same HLL framework. Both numerical analysis and computational results clarify that the modified flux function introduces no negative effects on resolving contact discontinuities and the boundary layer, which is critical for hypersonic viscous flow computations especially the heating issue. Numerical results that are obtained for various test cases indicate that both all-speed HLL-type schemes have a good performance in terms of accuracy and robustness for a broad spectrum of Mach numbers and they can also be used as reliable tools for a practical hypersonic heating problem involving complex geometries. The developed all-speed HLL-type solvers can be also applied to heat flux prediction on unstructured grids with multidimensional reconstruction method, which will be considered in the further investigation.

\section*{Acknowledgement}
\label{S:7}
This work was supported by the National Natural Science Foundation of China (Grant 11472004) and the Foundation of Innovation of National University of Defense Technology (Grant B150106).

\bibliography{manuscript}

\end{document}